%% file: sn-article.tex
\documentclass[sn-mathphys]{sn-jnl}% Math and Physical Sciences Reference Style

\usepackage{adjustbox,amssymb,pifont}
\newcommand{\xmark}{\ding{55}}
\usepackage[justification=centering]{caption}
\usepackage{tabularx}
\usepackage{graphicx} % Required for including images
\usepackage{subcaption} % Required for subfigures
\usepackage{bbm}
\usepackage{tablefootnote}
\usepackage{placeins}
\jyear{2023}%

\theoremstyle{thmstyleone}%
\newtheorem{theorem}{Theorem}%  meant for continuous numbers
\newtheorem{proposition}[theorem]{Proposition}% 

\theoremstyle{thmstyletwo}%
\newtheorem{example}{Example}%
\newtheorem{remark}{Remark}%

\theoremstyle{thmstylethree}%
\newtheorem{definition}{Definition}%

\begin{document}

\title[Article Title]{Generalizing Multimorbidity Models Across Countries: \\ A Comparative Study of Austria and Denmark}


\author*[1,2,3]{\fnm{Johanna} \sur{Einsiedler}}\email{johanna.einsiedler@sodas.ku.dk}

\author[4]{\fnm{Katharina} \sur{Ledebur}}\email{ledebur@csh.ac.at}

%\author[2]{\fnm{Jolien} \sur{Cremers}}\email{joc@dst.dk}

\author[4,5]{\fnm{Peter} \sur{Klimek}}\email{klimek@csh.ac.at}
\author[2,6,7]{\fnm{Laust Hvas} \sur{Mortensen}}\email{lhm@rff.dk}
\small
\affil*[1]{\footnotesize\orgdiv{Center for Social Data Science}, \orgname{University of Copenhagen}, \orgaddress{\street{Oster Farimagsgade 5}, \city{Copenhagen}, \postcode{1353}, \country{Denmark}}}

\affil[2]{\footnotesize\orgdiv{Data Science Lab}, \orgname{Statistics Denmark}, \orgaddress{\street{Sejrøgade 11}, \city{Copenhagen}, \postcode{2100}, \country{Denmark}}}
\affil[3]
{\footnotesize\orgdiv{Digital Humanities Lab}, \orgname{University of Basel}, \orgaddress{\street{Allschwilerstrasse 10}, \city{Basel}, \postcode{4055}, \country{Switzerland}}}
\affil[4]{\footnotesize\orgname{Complexity Science Hub Vienna}, \orgaddress{\street{Josefstädter Strasse 39}, \city{Vienna}, \postcode{1080},  \country{Austria}}}
\affil[5]{\footnotesize\orgdiv{Section for Science of Complex Systems, CeMSIIS}, \orgname{Medical
University of Vienna},
\orgaddress{Spitalgasse 23}, \city{Vienna}, \postcode{1090}, \country{Austria}}
\affil[6]{\footnotesize\orgdiv{Department of Public Health}, \orgname{University of Copenhagen}, \orgaddress{\street{Oster Farimagsgade 5}, \city{Copenhagen}, \postcode{1353}, \country{Denmark}}}
\affil[7]{\footnotesize\orgname{ROCKWOOL Foundation}, \orgaddress{\street{Ny Kongensgade 6}, \city{Copenhagen}, \postcode{1472}, \country{Denmark}}}


\abstract{%In aging societies, multimorbidity is a major challenge for individuals, health care providers, and society as a whole.

Chronic diseases frequently co-occur in patterns that are unlikely to arise by chance, a phenomenon known as multimorbidity. This growing challenge for patients and healthcare systems is amplified by demographic aging and the rising burden of chronic conditions. However, our understanding of how individuals transition from a disease-free state to accumulating diseases as they age is limited. Recently, data-driven methods have been developed to characterize morbidity trajectories using electronic health records; however, their generalizability across healthcare settings remains largely unexplored. In this paper, we conduct a cross-country validation of a data-driven multimorbidity trajectory model using population-wide health data from Denmark and Austria. Despite considerable differences in healthcare organization, we observe a high degree of similarity in disease cluster structures. The Adjusted Rand Index (0.998) and the Normalized Mutual Information (0.88) both indicate strong alignment between the two clusterings. These findings suggest that multimorbidity trajectories are shaped by robust, shared biological and epidemiological mechanisms that transcend national healthcare contexts.}

\maketitle

\include{introduction}

\include{results}

\include{discussion}

\include{methods.tex}

\appendix

\include{appendix}

\bibliography{sn-bibliography}% common bib file


\end{document}

%% file: introduction.tex
\section{Introduction}
%Over the past century, rising life expectancy has transformed many Western societies into rapidly aging populations. This has been a great success, but not without its costs. As people live longer and healthcare improves, multimorbidity--the co-occurrence of multiple chronic diseases--has become increasingly common worldwide\citep{Chowdhury2023}. 
Chronic diseases rarely occur in isolation. They tend to cluster and co-occur more frequently than would be expected by chance, reflecting shared risk factors, biological interactions, and social determinants of health. This accumulation of conditions, known as multimorbidity, poses a significant challenge to individuals, healthcare providers, and society as a whole \citep{Chowdhury2023}. The increasing prevalence of multimorbidity worldwide is a growing concern as populations age and life expectancy increases.
%Multimorbidity poses a major challenge to individuals, healthcare providers and the society as a whole. 
However, 
little is known about how individuals progress from a disease-free state to multiple coexisting conditions. Most research to date has focused either on single diseases or on simply counting co-existing diseases \citep{Marengoni2008,Line056135, Van_den_Bussche2011,Ho2021}. 

An alternative approach is to study disease trajectories, the sequences of diseases that unfold over the life course. %Consequently, methods for constructing and analyzing multimorbidity trajectories remain underdeveloped \citep{Jorgensen2024}.
Unlike static counts, trajectory-based methods capture the ordering, timing, and combinations of diseases. This provides insight into the underlying dynamics and patterns of multimorbidity.

Recent advances in statistical and machine learning methods, particularly those leveraging electronic health records (EHRs), have provided new opportunities to model disease trajectories \citep{Jorgensen2024, Chmiel_2014, Jensen2014, Jeong2017,Fotouhi2018,Roque2011,Giannoula2018}. However, nearly all applications have relied on data from a single country, raising questions about the robustness and generalizability of the identified patterns as reported trajectories may reflect not only underlying disease processes but also country-specific healthcare practices or data structures.
While some cross-national comparisons of multimorbidity exist \citep{Garin2016Global,Arokiasamy2015Impact}, these largely rely on survey data.

Because EHRs contain sensitive, personally identifiable information, sharing data across research groups is difficult, so comparative studies in this field are rare. As highlighted in a recent review, “very few studies have validated disease trajectories with populations from other countries and in these few cases, they often replicate their own disease trajectories based on summary data” \cite{Jorgensen2024}. In this study, we address this gap by conducting a harmonized cross-country comparison of multimorbidity trajectories based on hierarchical clustering applied to individual-level health records from Austria and Denmark.

Austria (AT) and Denmark (DK) are both high-income, mixed-economy countries in the EU with populations of 8.51 mio. (AT) \citep{ATpop}  and 5.63 mio. (DK) \citep{DSTpop} in our reference year 2014. The age structure, life expectancy (AT: 81.2 years \citep{ATlife}, DK: 80.5 years \citep{DKlife}; 2014) and burden of disease is similar \citep{GBD2020}. Austria's healthcare system is based on the Bismarck model, which is characterized by mandatory social health insurance and a significant role for employment-based funds. Notably, Austria has a dual financing system with insurance-based financing in the outpatient sector and tax-based financing in the inpatient sector \citep{bmasgk2019austrian}. Denmark's healthcare system follows the Beveridge model, with tax-based funding and centralized control.
Health care spending was 10.4\% (AT) and 10.6\%  (DK) of GDP in 2015 \citep{oecd_who_2015}.

%Health care spending was 12.1\% (AT) and 10,8\% (DK) of GDP in 2021\citep{DK2023, AustriaCountryHealth2023}.
In both countries, about 15\% of health care spending was out-of-pocket. The primary health care sector plays an important role as gatekeepers to specialists and hospitals in both countries. The division of labor between hospitals and non-hospital providers is substantially different, which affects the numbers and characteristics of patients treated at hospitals  \citep{Reibling2012}: Austria emphasizes inpatient care (735 hospital beds per 100,000 inhabitants; 2015 \cite{eurostat2025beds}), while Denmark emphasizes outpatient and community-based service delivery (248 hospital beds per 100,000 inhabitants; 2015 \cite{eurostat2025beds}).

In the present study we assess the extent to which disease trajectory patterns replicate across these two distinct healthcare contexts. By doing so, we provide new evidence on the robustness and generalizability of data-driven approaches for studying multimorbidity, offering insights into both the biological and health system determinants of disease progression.

% 541 phy and 1060 nurses per 100k in AT https://iris.who.int/bitstream/handle/10665/379901/9789289014335-eng.pdf?sequence=1
% 438 phys and 1024 nurses per 100k in DK https://iris.who.int/bitstream/handle/10665/376807/9789289059626-eng.pdf?sequence=1

% 672 hospital beds for AT and 248 hospital beds for DK (per 100 k ) https://ec.europa.eu/eurostat/databrowser/view/hlth_rs_bdsrg2/default/table?lang=en&category=hlth.hlth_care.hlth_res.hlth_facil

%% file: results.tex
\section{Results}

\subsection{Multimorbidity Patterns in Austria and Denmark}

\begin{figure}[h!]
\centering
\includegraphics[width=\linewidth]{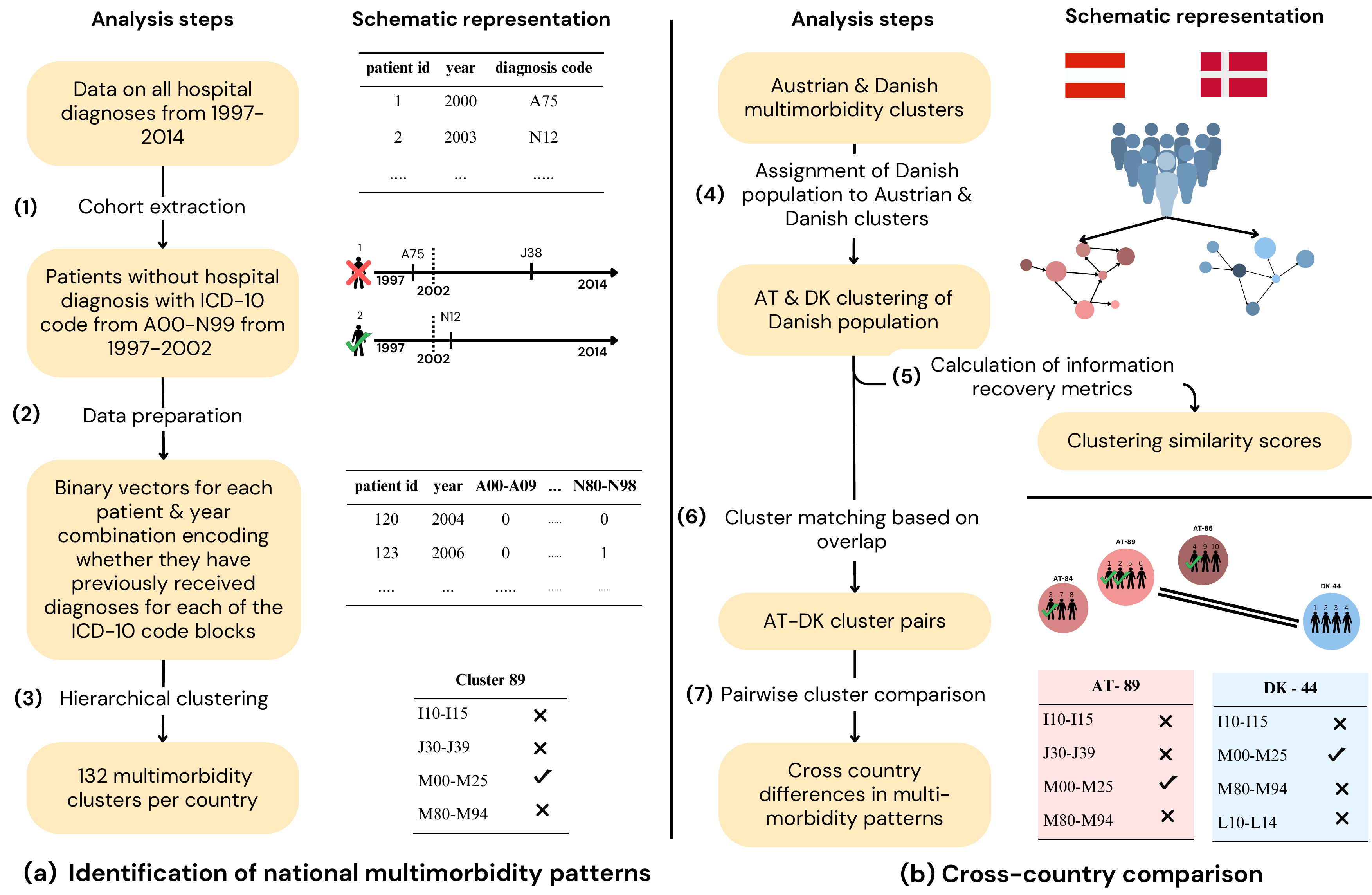}

\caption{Workflow of the analysis presented in this article. Boxes represent data entities, arrows represent analysis steps. (a) Identification of national multimorbidity patterns. Patient-year health states were derived from national hospital registers, including all individuals with at least one in-patient hospital stay between 2003 and 2014 and no stay between 1997 and 2002. For each patient and year, binary vectors were constructed to indicate whether diagnoses had been recorded in each diagnose block, covering ICD-10 diagnoses A00–N99. Applying a hierarchical clustering algorithm to these health states identified 132 multimorbidity clusters per country. (b) Cross-country comparison. Danish patients were assigned to both Austrian and Danish clusters. Information recovery metrics were calculated, clusters were matched based on overlap, and pairwise cluster comparisons were performed to assess cross-country differences in multimorbidity patterns.}
\label{fig:analysis_flow}
\end{figure}

\begin{figure}
\centering
\includegraphics[width=\textwidth]{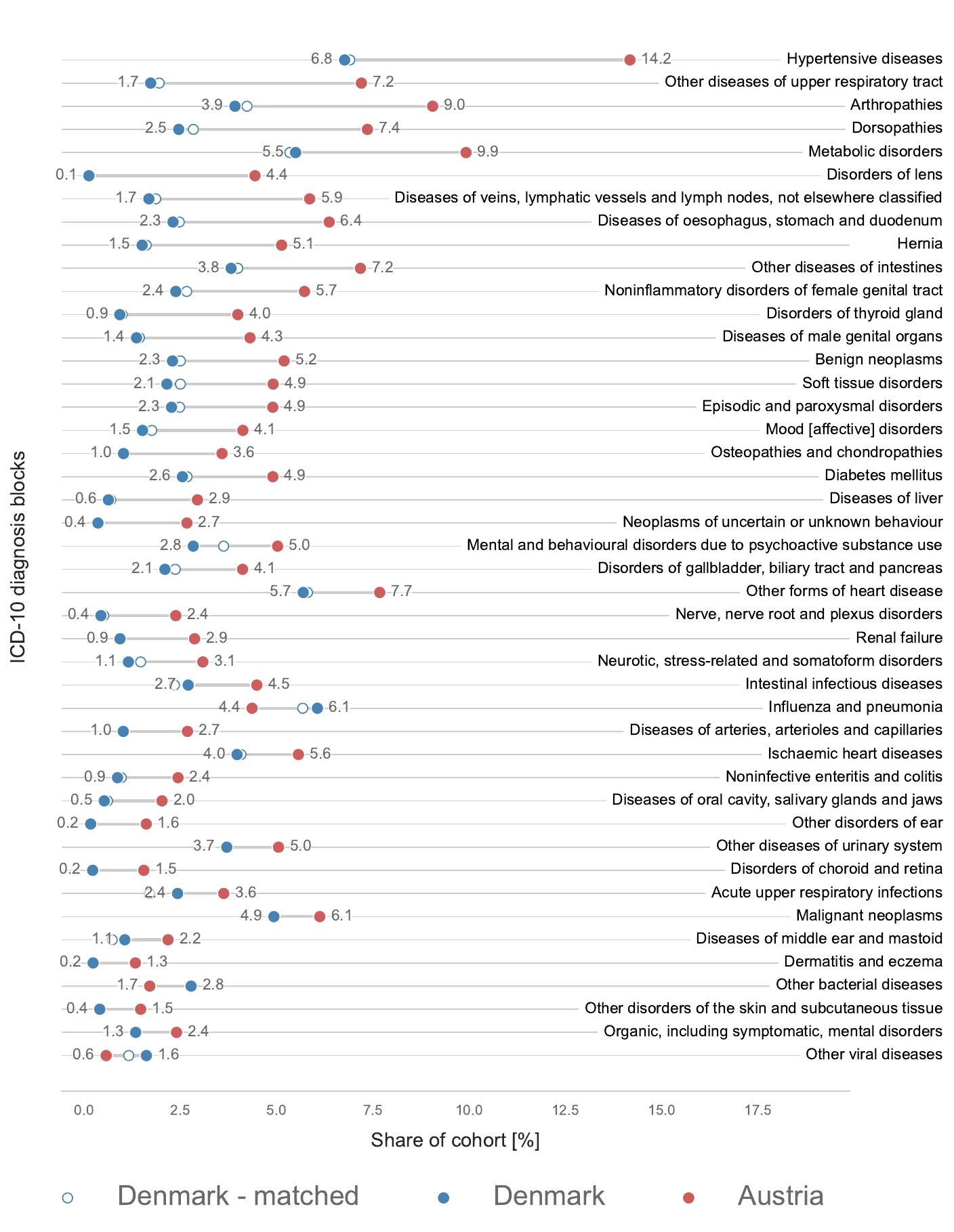}

\caption{Share of cohort that received a diagnosis from a specific diagnosis block within the observation period for the Austrian cohort, the Danish cohort and the age \& gender adjusted Danish cohort. Only diagnosis blocks are shown where the difference between the Austrian and the Danish sample exceeded 1pp.}
\label{fig:diag_comparison}
\end{figure}

Following the workflow illustrated in Figure \ref{fig:analysis_flow}, we analyze groups of individuals without any recorded in-patient hospital stays between 1997 to 2002 and at least one stay between 2003-2014. We use data from the Danish National Patient Register \cite{Schmidt2015} and the Austrian Ministry of Health's registry. 
We find notable differences in the types of diagnoses assigned in Denmark and Austria. Figure \ref{fig:diag_comparison} presents the share of individuals in each cohort who received a diagnosis code during the observation period, restricted to diagnosis blocks where the difference between countries exceeded 1 percentage point. The largest discrepancies are observed for hypertensive diseases (Austria: 14.2\%; Denmark: 6.8\%) and other diseases of the upper respiratory tract (Austria: 7.2\%; Denmark: 1.7\%). In general, as expected based on the different health system structures, most diagnoses are more frequently recorded in hospitals in Austria than in Denmark.

Adjusting the Danish sample to match the Austrian gender and age distribution reduces the differences for nearly all diagnosis blocks. The only exceptions are soft tissue disorders, intestinal infectious diseases, and breast disorders, where disparities persist.  Nonetheless, substantial differences in diagnoses patterns remain, suggesting that these patterns cannot be fully explained by demographic composition. Given that official statistics report broadly comparable disease prevalence in the two countries, a plausible explanation is that certain conditions are more often diagnosed outside the hospital setting in Denmark.

%Absolute counts of most frequently occurring diagnoses as well as diagnosis pairs are reported in Appendix \ref{app_B:most_frequent}.

\subsection{Comparing Austrian and Danish Multimorbidity Clusters}

For identifying multimorbidity clusters, we follow \citep{Haug2020} and define the health state of an individual in a given year as the collection of all disease categories (ICD-10 codes A00–N99) in which they had previously received a diagnosis. By applying the hierarchical clustering algorithm DIVCLUS-T to these patient-year health states, we identify 132 multimorbidity clusters. Each cluster represents a distinct multimorbidity pattern of co-occurring conditions, defined by characteristic inclusion and exclusion criteria.

%\cite{Haug2020} identified 132 multimorbidity patterns by applying a hierarchical clustering algorithm to patient hospital diagnosis histories in Austria. Each cluster is characterized by a set of diseases that all individuals within the cluster share (inclusion criteria) and by a set of diseases from which they must be free (exclusion criteria).
As a first step, we attempted to replicate the clusters identified by \citep{Haug2020} using the Austrian data.
Results were not identical but had only very minor differences, potentially caused by different computing environments. We report all differences in detail in Appendix \ref{app:replication}. In the next step, we applied the same clustering procedure to the Danish cohort. 
\subsubsection{Summary Statistics}
\begin{figure}[htbp]
    \centering
    \captionsetup{aboveskip=0pt, belowskip=10pt}
    
    % --- Left subfigure ---
    \begin{subfigure}[t]{0.48\textwidth}
        \centering
        \includegraphics[width=\textwidth]{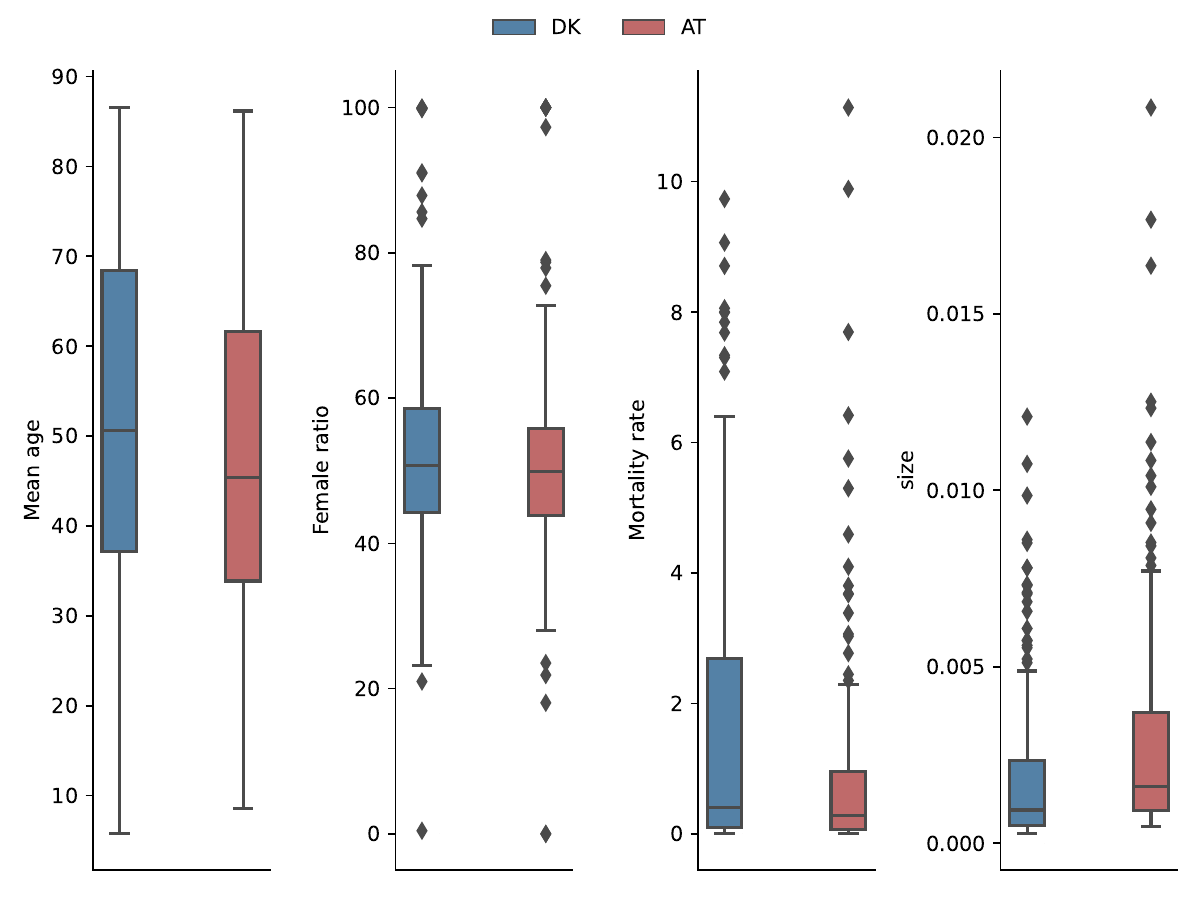}
        \caption{Comparison of distributions of sizes, mean age, percentage of females, and mortality percentages in the Austrian and the Danish clustering.}
        \label{fig:cluster_boxplot_comparison}
    \end{subfigure}
    \hfill
    % --- Right subfigure ---
    \begin{subfigure}[t]{0.45\textwidth}
        \centering
        \includegraphics[width=\textwidth]{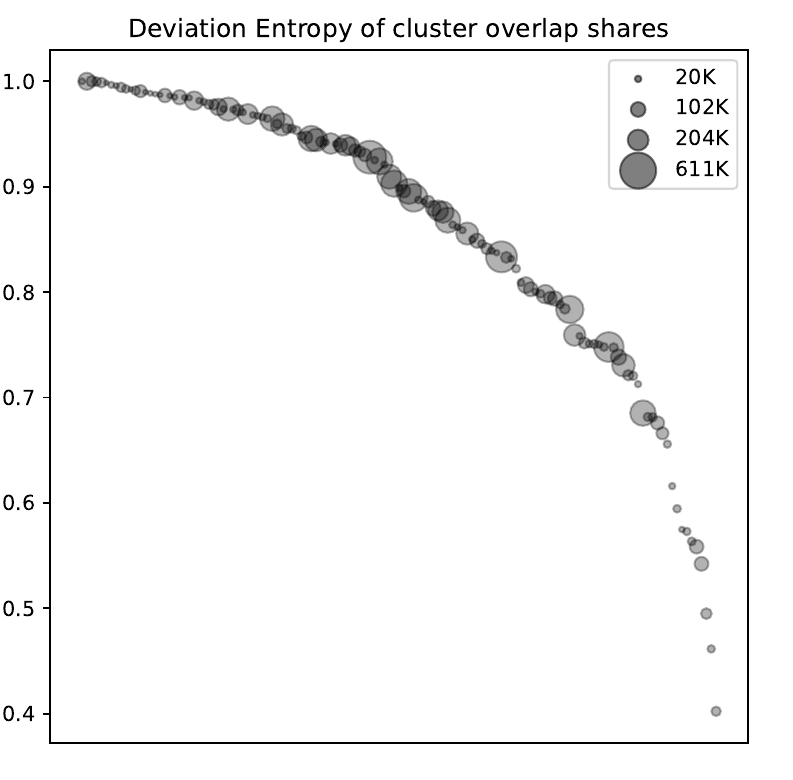}
        \caption{Deviation entropy of cluster overlap shares ranked from highest to lowest. Marker size encodes the size of the respective Danish cluster.}
        \label{fig:deviation_entropy}
    \end{subfigure}
    
    \caption{Comparison between Austrian and Danish cluster characteristics and overlap patterns.}
    \label{fig:cluster_distributions}
\end{figure}

As depicted in Figure \ref{fig:cluster_boxplot_comparison}, overall we find the distribution of the sizes of the clusters in both countries to be fairly similar. However, we do observe a a slightly larger share of comparatively smaller clusters in Denmark. Per definition, each clustering has one cluster with all observations   that have not yet received any diagnoses. This ``zero-cluster'' is the largest cluster in both countries. Apart from the zero-cluster, the three largest Danish clusters are characterized by single inclusion criteria: \emph{Arthropathies} (Cluster 83, representing 1.2\% of observations), \emph{Malignant neoplasms} (Cluster 92, 1.0\%), and \emph{Influenza pneumonia} (Cluster 67, 0.8\%). In Austria the largest clusters are defined by the inclusion criteria \emph{Other diseases of upper respiratory tract} (Cluster 92, 2.1\% of observations), \emph{Arthropathies} (Cluster 85, 2.8\% of observations) and \emph{Noninflammatory disorders of female genital tract} (Cluster 81, 1.6\% of observations).
We observe a larger share of Danish clusters with a mean age below twenty, as well as a higher number of clusters with mean age around 50 and around 70, reflecting also the differences in demographics between the two samples (see Appendix \ref{app:sample_diff}). Regarding the shares of males and females in the clusters the distribution across clusters is fairly similar in both clustering structures. For mortality rates, we also observe broadly similar distributions, although Austria shows a slightly heavier left tail. This likely reflects the greater extent to which less severe cases are managed in outpatient care in Denmark.
The mean number of inclusion criteria in the Austrian clustering is 1.7, compared to 1.53 in Denmark. For exclusion criteria, we have 18 on average in Austria and 22 in Denmark (see Appendix \ref{app_c} for more details).

The cluster that includes \emph{Malignent neoplasms} and \emph{Influenza pneumonia} (Cluster 107) has the overall highest hospital mortality rate (9.7\%) in Denmark. Among the 10 clusters with the highest mortality rate in Denmark, 7 have \emph{Malignent neoplasms} as an inclusion criterion. \emph{Hypertensive diseases} and \emph{Heart diseases}, which were relevant for the highest mortality rate cluster in Austria, only occur in 3 out of these 10.
All inclusion and exclusion criteria for all the obtained clusters are listed in Appendix \ref{app:cluster_criteria}.

%The clusters with the highest mean age in Denmark are characterized by \emph{Metabolic disorders}, \emph{Hypertensive diseases} and \emph{Other diseases of urinary system} (Cluster 127, $m=83 years$); \emph{Ischaemic heart diseases}, \emph{ Ohter forms of heart diseases} and \emph{Influenza and pneumonia} (Cluster 110, $m=82$ years) as well as  \emph{Hyptertensive disease}, \emph{Other forms of heart disease} and \emph{Influenza and pneumonia} (Cluster 124, $m=80$ years). In Austria those are 

%Multimorbidity clusters derived from Austrian data have an average of 1.7 inclusion criteria and 19.0 exclusion criteria. The clusters of the Danish population have an average of 1.6 inclusion criteria and 21.6 exclusion criteria. The distribution is shown in Figure \ref{fig:criteria_dist}.

\subsubsection{Cluster Overlap and similarity measures}

To compare the cluster structures between the Austrian and Danish populations, we assigned the Danish population to both the Danish clusters and the Austrian clusters, and then analyzed the population overlap between clusters. Figure \ref{fig:deviation_entropy} illustrates the distribution of the cluster overlap deviation entropy, which quantifies how concentrated or dispersed the members of a given Danish cluster are across Austrian clusters. The majority of clusters has a deviation entropy higher than 0.8, meaning that most individuals within these Danish clusters are consistently assigned to a single Austrian cluster. Conversely, the lower values at the tail of the distribution reflect greater dispersion, suggesting that the same Danish cluster’s members are distributed across multiple Austrian clusters, indicative of structural divergence between the two clustering solutions.

%illustrates the proportion of each Danish cluster (y-axis) that is allocated to each Austrian cluster (x-axis). A line containing many highlighted boxes indicates that the respective Danish cluster is distributed across multiple Austrian clusters. On average, the population of a Danish cluster is spread across 17 Austrian clusters. Nevertheless, 87 out of 132 Danish clusters have more than 80\% of their population overlapping with a single Austrian cluster.

Generally, our results reveal a high degree of correspondence between the two models. For 87 Danish clusters, more than 80\% of their members are assigned to a single Austrian cluster. When considering the two Austrian clusters with the largest overlaps, 104 Danish clusters show a combined overlap exceeding 80 \%. Overall, 85.7\% of all observations (excluding those in the zero cluster) are assigned to the Austrian cluster with the highest overlap, while an additional 4.4\% are assigned to the Austrian cluster with the second-highest overlap.

Based on the assignment of Danish individuals to both clustering solutions, we computed standard measures of cluster similarity: The resulting normalized mutual information between the Danish and Austrian clustering, is 0.88. The adjusted rand index is 0.998. It holds for both measures, that a value close to 1 indicates high similarity. The reported scores thus indicate a high degree of similarity of the cluster structures. 

%If we match each Danish cluster to the Austrian cluster with the highest overlap in assigned observations, 7 clusters have an identical population, 60 clusters have an overlap of over 90\%. The Danish cluster with the lowest overlap with a single Austrian cluster has 26\% overlap. Overall, 7\% of the whole population are not allocated to the best matching cluster in the Austrian clustering of their respective Danish cluster.

% \subsection{Inclusion criteria and diagnosis frequency}
% The idea behind the clustering is to capture frequent multimorbidity patterns. Thus, we observe a high correlation of the frequency of diagnosis of a particular disease and its occurrence as an inclusion criterion in the clustering.

% For the Danish clusters, see Figure \ref{fig:hist_diag_incl_DK}, Arthropathies is the most diagnosed disease and simultaneously an inclusion criterion in 18 clusters. The second most frequent inclusion criterion is Soft Tissue Disorders, which occurs over 3.5 million times in the Danish dataset. Interestingly, Neurotic, stress-related and somatoform disorders occurs as inclusion criterion comparatively often while being rather rare in the population, indicating that it is a part of very distinctive multimorbidity patterns.

% For the Austrian clusters, see Figure \label{fig:hist_diag_incl_AT}, hypertensive diseases is by far the most common diagnosis and also occurs most often as an inclusion criterion in the clustering (30 times). Also metabolic disorders play a large role in characterizing common multimorbidity patterns in Austria.

\subsection{Cluster-specific cross-national comparisons}
To assess the similarities and differences at a clinically meaningful disease-specific level, we compare clusters within both countries that have the diagnosis of widely studied diseases as inclusion criteria: Diabetes mellitus (E10-E14), ischaemic heart diseases (I20-I25), chronic lower respiratory diseases (J40-J47), cerebrovascular diseases (I60-I69), malignant neoplasms (C00-C97) and affective disorders (F30-F39). 
We find comorbidity patterns and trajectories into and out off these clusters to be similar between Austria and Denmark with only minor differences: %e.g. a higher prevalence of influenza-induced pneumonia as a comorbidity in the Austrian clusters that include chronic lower respiratory diseases or differences in the relative size of clusters. 
For diabetes, the most common trajectories in both countries begin with hypertension and progress to include metabolic disorders, matching well-established patterns of cardiometabolic comorbidity. For ischemic heart disease, individuals typically transition from clusters capturing ischemic or other heart diseases to clusters combining both, with nearly identical trajectories in both countries. For malignant neoplasms, most patients in both countries transition directly from a healthy state to a cluster defined solely by neoplasms. Frequent next steps involve intestinal or blood-related conditions. For a more detailed comparison of the country specific trajectories, see Appendix \ref{app:selected-diseases}.   

\subsection{Clusters with a low overlap}

At the same time, some differences in cluster characteristics remain. We observe the lowest overlap with a single Austrian cluster (20-33\%) for Danish clusters 100, 114, and 115.  Danish cluster 100, which is defined by metabolic disorders, is split across Austrian clusters that focus on renal, intestinal, and stress-related conditions. This suggests that patients with metabolic disorders in Austria are more often admitted with co-occurring somatic or psychosomatic diagnoses than patients in Denmark, where such combinations may be less frequently captured in hospital records. Similarly, Danish clusters 114 and 115 group chronic lower respiratory diseases with influenza and pneumonia, but exclude hypertensive disease. These clusters show only partial overlap with Austrian cluster 48. While cluster 48 contains similar cardiac and respiratory conditions, it does not exclude hypertension and likely reflects broader multimorbidity profiles.

%% file: discussion.tex
\section{Discussion}
 
This study emphasizes the transferability of multimorbidity trajectory models across different national health systems. Despite clear differences in how care is organized, financed, and accessed in Austria and Denmark, the two countries exhibit a high degree of similarity in disease cluster structures. Both the Adjusted Rand Index (0.998) and the Normalized Mutual Information (0.88) indicate strong alignment between the two cluster structures. Moreover, for common chronic conditions such as diabetes, ischemic heart disease, and malignant neoplasms, individuals follow similar health trajectories and are assigned to clusters with nearly identical inclusion criteria. These results suggest that clustering methods primarily capture underlying disease patterns rather than country-specific artifacts in healthcare data.

Given the institutional differences between the Austrian and Danish healthcare systems, the stability of the clustering results is notable. Austria permits direct access to specialists and hospitals and has a higher density of hospital beds and physicians per capita~\cite{AustriaCountryHealth2023}. In contrast, Denmark relies on a strict gatekeeping model through general practitioners and has a more centralized system with fewer inpatient resources~\cite{Reibling2012}. Despite these differences in access to and delivery of care, clustering produces similar outputs in both settings. These results strengthen the case for applying trajectory models in cross-national studies, as these methods appear to reflect disease progression rather than artifacts of the health systems.

Trajectory analyses of diabetes, ischemic heart disease, and malignant neoplasms reveal similar and clinically plausible sequences in individuals in both Austria and Denmark. These consistent patterns demonstrate that clustering methods can reliably capture meaningful disease progressions across different healthcare settings. However, some differences remain, specifically regarding Danish individuals with metabolic disorders or chronic lower respiratory diseases. These patterns could result from Austria's more open access to hospital care and its tendency toward more comprehensive diagnostic coding per admission. In contrast, Denmark's gatekeeping model may lead to more selective hospital admissions and a narrower diagnostic focus. Together, these differences could contribute to the observed misalignment between certain clusters.

%LIMITATIONS
This study also comes with several limitations. First of all, the age difference in our samples likely contributes to relative differences in the prevalence of age-specific diseases. This may have influenced the clustering results, as certain trajectories could appear more or less prominent simply due to cohort composition. While our comparison of disease prevalence in a Danish sub-sample that mirrors the Austrian age and gender composition, indicates that age is not the main driving factor, we are not able to ultimately rule out that it influenced our conclusions.
Further, data quality and completeness of Administrative health records depends on coding accuracy, consistent recording and coding norms within institutions. Errors, omissions or differences in coding diseases could thus bias cluster formation and trajectory identification.
Finally, both Austria and Denmark are high-income European countries with relatively comprehensive healthcare systems. Findings may not generalize to countries with different disease burdens, resource constraints, or healthcare access patterns (e.g., low- and middle-income countries).

% billing purposes, difference in age structur

% CONCLUSION
Future studies should therefore test these models in non-European and lower-income countries, where disease burdens, access barriers, and data quality tend to differ substantially. Studies should also examine how cluster structures vary by age, over time, or in response to events such as pandemics. Identifying where models are applicable and where they diverge will help refine their use in prevention and care planning.

Nonetheless, our findings indicate that clustering methods offer a scalable approach to identifying stable disease patterns across populations. Their robustness to differences in national health systems makes them a promising tool for comparative health research and for developing system-independent approaches to managing multimorbidity.

% suggest that pathophysiological co-occurrence patterns dominate over national differences - BUT: social determinants of health, common risk exposures (smoking, diet)

%% file: methods.tex
\section{Methods}

\begin{comment}
for writing the intro and methods, I think we agreed to describe the study symmetrically in terms of populations and clusterings. Hence, data and clustering methods should ideally be introduced without referring to any nationality (ie., not saying "we took Austrian clusters and compared them to assignments of DK patients, etc.). There also needs to be some more detailed description of the study design, maybe a flow chart in which we give a graphical overview of the analysis steps? "Patient overlap" should also be described as a similarity indicator.

\end{comment}

\subsection{Identification of national multimorbidity patterns}

\textbf{Cohort Extraction.} 
Following \citep{Haug2020}, we define our cohort as all individuals who (1) have been assigned at least one ICD-10 diagnosis code between A00 and N99 in a hospital between 1997 and 2014, (2) have not been diagnosed with any diseases with codes A00 to N99 between 1992 and (including) 2002 and (3) were born before the beginning of the year 2013. Thereby, we obtain a cohort that can be considered as "healthy", at the beginning of the analysis period in 2003.  We restrict our analysis to all diagnoses from A00 to N99 that these individuals received between 2003 and 2014. As defined by the WHO  \cite{ICD-10}, we group those into 131 disease blocks (see Figure \ref{fig:analysis_flow}a, Step 1).
 As reported in Table \ref{tab:overview} the resulting datasets consist of 42,093,844  observations in Denmark and 66,466,543 observations in Austria. In the Danish dataset out of these 71.8\% are healthy observations, i.e. individuals who haven't had any diagnoses up until at least that specific year (AT: 59.4\%). This means that individuals in the Danish cohort have their first hospital contact (within the observation period) comparatively late. Partially, this is also the cause for the lower average number of first-time diagnoses per individual (1.28 in Denmark; 2.56 in Austria). The pattern also holds true when looking at different age cohorts separately (see Appendix \ref{app_B:density} for more details). The resulting cohorts consist of 5,112,811 ($ \sim 60\%$ of population in 2014) and 3,237,988 ($\sim 58\%$ of population in 2014) individuals from Austria and Denmark, respectively. The sex ratio is relatively similar, with the share of females being 53.3\% in Austria and 55.7\% in Denmark. When it comes to age, however, we find large differences: At the end of the observation period, i.e. 2014, the mean age in the Austrian sample is 56 years, while it is 41 years in Denmark. This is largely caused by a relatively large number of individuals born after the year 2000 that are included in the Danish sample. We find this to be consistent with current national reports on hospital admissions and population demographics (see Appendix \ref{app:sample_diff} for a more detailed report). 
 %Due to the large differences in age, we also create a matched sub-sample of the Danish cohort that mirrors the demographics (i.e. 5-year age groups and gender) of the Austrian cohort. 
\renewcommand{\arraystretch}{1.5} % 1.5x row height

\begin{table}[h!]
\centering
\caption{Data overview}
\label{tab:overview}

\begin{tabular}{p{5cm}cc}
\hline

 & Austrian data & Danish data \\ \cline{2-3}
 
 & \multicolumn{2}{c}{Initial dataset size (1997-2014)} \\ \cline{2-3}
A-Z diagnosis codes within hospital    & ~120,000,000      & 25,675,195   \\ 
Individuals diagnosed with at least one diagnosis code from A-Z in hospital     & ~9,000,000      &  4,795,961     \\ 
Inpatient hospital stays with a primary or secondary diagnosis code from A-Z & 45,000,000 & 14,185,892\\

 & \multicolumn{2}{c}{Cohort characteristics} \\ \cline{2-3}
Size    & 5,112,811     &  3,237,988 \\ 
Sex ratio - \% female & 53.3 & 55.7 \\ 
Average age in years & 56 & 40.5 \\ 

 & \multicolumn{2}{c}{Dataset characteristics} \\ \cline{2-3}
Observations &  66,466,543 & 42,093,844 \\
\% observations without diagnoses & 59.4 & 71.8 \\
Number of first-time diagnoses & 13,121,008  & 4,135,271\\ 
Average number of first-time diagnoses per individual & 2.56  & 1.28 \\
\hline
\end{tabular}
\end{table}

\begin{comment}
%this is 45,000,000 (number of hospital stays as written in the original paper) * 2.65 (i.e average number of diagnoses assigned per stay as written in the paper)
\end{comment}

\textbf{Dataset Construction}
For each hospital stay, the datasets contain the main diagnosis in the form of level-3 codes from the International Classification of Disease (ICD-10), the year of diagnosis, the sex of the patient (male/female) and in the Danish (Austrian) case the patient age (5-year age interval group). Accordingly, we define an individual's health state for each year as the set of all diagnosis blocks they have received diagnoses from so far. If an individual receives the same diagnosis more than once within the same year, this is only recorded once.
Thus, if an individual is diagnosed with a disease once, it will be recorded in their health state in the year they have been diagnosed as well as every succeeding year. Based on this we construct the main analysis dataset which consists of health states for each year in the observation period (2003 to 2014) and each individual in the cohort (see Figure \ref{fig:analysis_flow}a, Step 2).

\textbf{Hierarchical clustering} Next, we use DIVCLUST-T \cite{CHAVENT2007687}, a divisive clustering algorithm to group the recorded health states into $K=132$ clusters, representing common multimorbidity patterns (see Figure \ref{fig:analysis_flow}a, Step 3). To ensure comparability we follow the exact procedure by \cite{Haug2020}. Each cluster is defined by a list of diseases that all individuals in this cluster have been diagnosed with (inclusion criteria) as well as a list of diseases they must not have been diagnosed with (exclusion criteria). Individuals within a cluster can differ with respect to diseases neither listed as an inclusion nor as an exclusion criterion. Details on the demographics of these clusters in comparison to the Austrian clustering can be found in Appendix \ref{app_c}.

\subsection{Cross-country comparison}

To compare the similarity of the Danish and the Austrian multimorbidity patterns, we assign all Danish observations to the clusters derived from Austrian data. Thereby, we essentially create two different schemes of categorizing each Danish individual for each year. Based on these assignments we can then calculate common cluster similarity measures and compare clusters with a similar population from each clustering scheme.
\subsubsection{Similarity measures}
To get an estimate of the degree of similarity between the clusters, we calculate the normalized mutual information score (NMI) and the adjusted rand index (ARI), two information recovery metrics commonly used in machine learning  \cite{Emmons2016}.

\textbf{Normalized Mutual Information Score.}

The Normalized Mutual Information score  (NMI) is a measure to quantify the degree of shared information between two data distributions. The NMI between the two clusterings, $C^{AT}$ and $C^{DK}$ can be defined as \citep{DAS2023135}:

$$NMI(C^{AT},C^{DK}) = \frac{2 \times I(C^{AT}, C^{DK})}{H(C^{AT})+H(C^{DK})}$$
 where $I(C^{AT},C^{DK})$ is the mutual information shared between $C^{AT}$ and $C^{DK}$. Further, $H(C^{AT})$ and $H(C^{DK})$ are the entropy of $C^{AT}$ and $C^{DK}$, respectively. The NMI can take on values between 0 (no mutual information) and 1 (perfect correlation).\\

\textbf{Adjusted Rand Index.}

The Adjusted Rand Index (ARI) is a measure of the overlap between two clusterings. It is calculated by counting all pairs of samples that are assigned to the same vs. different clusters. It is subsequently adjusted for random chance. The ARI ranges from -0.5 (discordant clusters) to 1 (identical clusters). An ARI of 0 means that the clusterings are independent of each other. A detailed explanation of the calculation procedure of the ARI can be found in \citep{TALBURT201163}.

\subsubsection{Cluster matching}
To identify relevant differences in the clusterings between one country and another, we calculate the pairwise overlap in population between each Danish and Austrian cluster. This enables the calculation of cluster overlap deviation entropy as a measure of dispersion of a Danish cluster's population across Austrian clusters.

\textbf{Population overlap.}

We define the population overlap as the share of observations in the Danish cluster $l$ that is present in the Austrian cluster $k$:

\begin{comment}
$$ Population Overlap_{lk} = \frac{ \sum_{i \in C^{DK}_l \land C^{AT}_k} i}{\sum_{i \in C^{DK}_l} i}$$

where $  \sum_{l = 1}^{132} \sum_{i \in C^{DK}_l} i$ is the sum of the population over all Danish clusters and thus equal to the size of the Danish dataset.
\end{comment}

$$ Population Overlap_{lk} = \frac{\sum_i \mathbbm{1}_{C_l^{DK} (i)\cap C_k^{AT}} (i)}{\sum_i \mathbbm{1}_{C_l^{DK}}(i)} $$

where $ \sum_{l=1}^{132} \sum_i \mathbbm{1}_{C_l^{DK}}(i)$ is the sum of the population over all Danish clusters and thus equal to the size of the Danish dataset.

\textbf{Deviation Entropy.}

The deviation entropy of cluster \( l \) is defined as the Shannon entropy of its distribution across clusters:

\[
H_l = - \sum_k p_{lk} \log(p_{lk}),
\]

where \( p_{lk} \) denotes the population overlap as defined above.

%% file: appendix.tex
\clearpage
\section{Sample differences}
\label{app:sample_diff}
As reported in the main manuscript, the birthyear distribution of the Danish study cohort is more heavily right tailed than the one of Austria.

\begin{figure}
  \begin{subfigure}{0.9\textwidth}
 \centering
    \includegraphics[width=\linewidth]{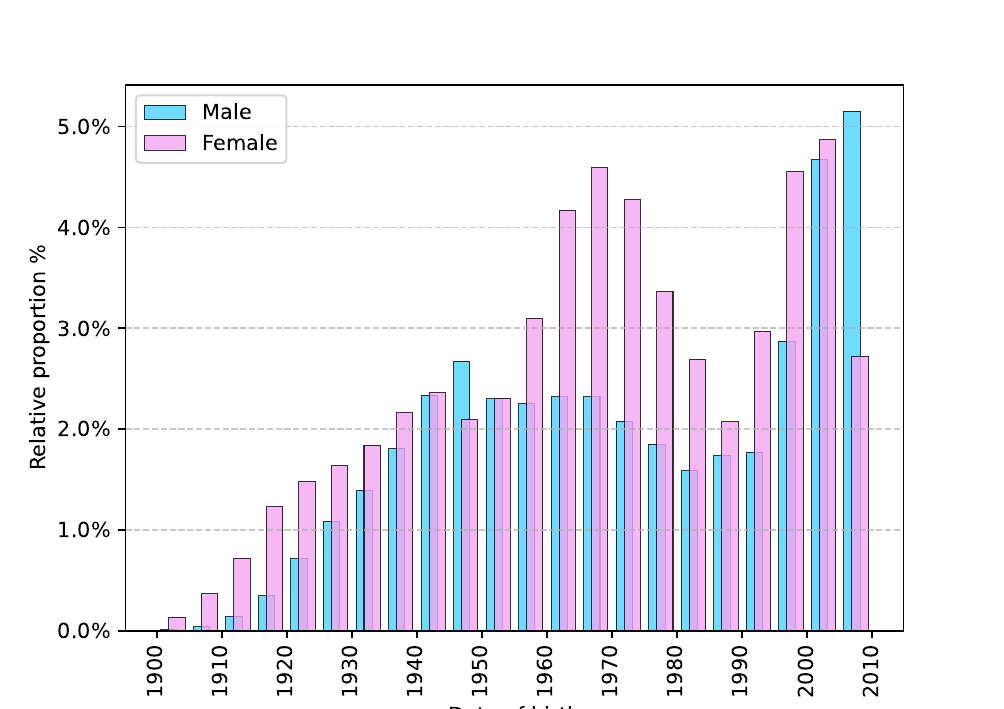}
\caption{Denmark}
    \label{fig:dist_age_sample_DK}

    \end{subfigure}
    \hfill
    \begin{subfigure}{0.9\textwidth}
        \centering
        \includegraphics[width=\linewidth]{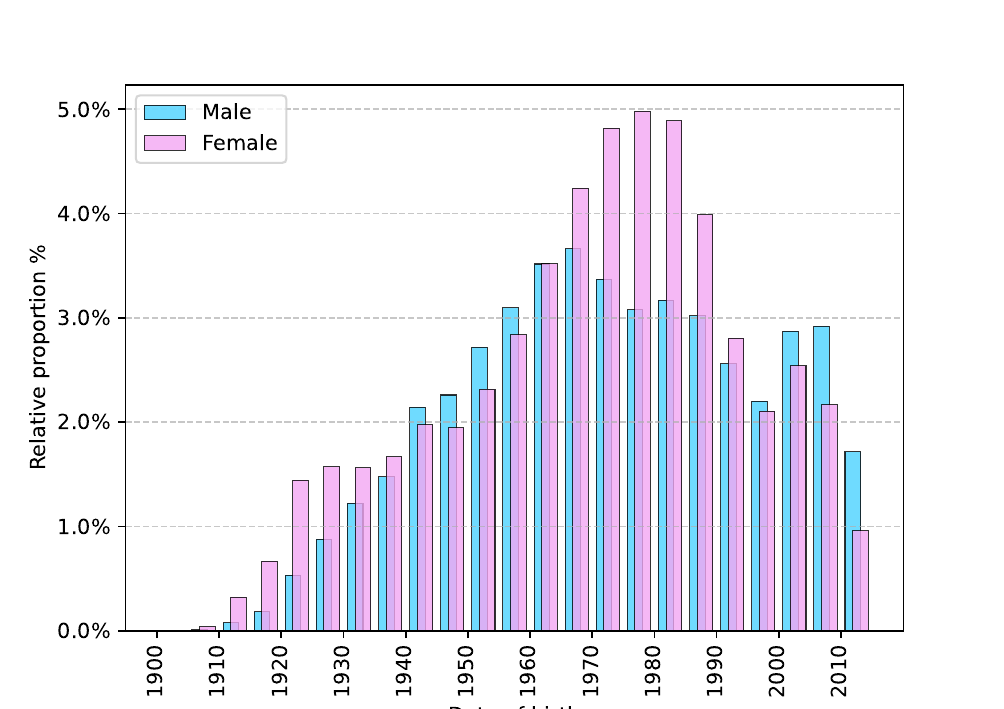}
        \caption{Austria}
        \label{fig:dist_age_sample_AT}
    \end{subfigure}
    \caption{Birthyear \& gender distribution of all individuals with (1)  at least one hospital diagnosis from ICD-10 blocks A00 to Z99 between 1997 and 2014 and (2) no diagnosis from A00 to N99 between 1997 and 2002}
   \label{fig:dist_age_sample}
\end{figure}

Figure \ref{fig:dist_age_sample} shows the different demographic structures of the Austrian and the Danish sample.

Since those individuals born after the year 200 are at maximum 14 years old at the end of the observation period in 2014, a potential explanation could be that those younger than 14 years old are more frequently admitted to hospitals in Denmark compared to Austria. Table \ref{tab:pop_stats} shows the numbers of hospitalized individuals grouped according to age at hospital admittance as well as the population demographics for both countries for the year 2014. These show very clearly that the percentage of under 14 year old individuals that are admitted to the hospital is much higher in Denmark (18.62\%) compared to Austria (5.92\%). This discrepancy is likely to explain the majority share of the differences in the sample demographics.

\begin{table}[h!]

\centering

\footnotesize
\begin{tabular}{lccc|ccc}
\hline
& \multicolumn{3}{c}{Austria}& \multicolumn{3}{c}{Denmark}\\
\midrule
Age & Hospitalized\tablefootnote{Statistics Austria: Hospital discharges 2014 by diagnosis, age, sex and federal province. \url{www.statistik.at/en/statistics/population-and-society/health/health-care-and-expenditure/inpatient-health-care-hospital-discharges}. Accessed 2024-12-14.} & Population\tablefootnote{Statistics Austria:  Bevölkerung nach Alter und Geschlecht seit 1869. \url{www.statistik.at/en/statistics/population-and-society/population/population-stock/population-by-age-/sex}. Accessed 2024-12-14} & \% & Hospitalized\tablefootnote{Statistics Denmark. BEFOLK2: Population 1. January by sex and age. \url{ww.statbank.dk/BEFOLK2}. Accessed 2024-12-14.} & Population\tablefootnote{Statistics Denmark. IND01: Admissions by region, diagnosis (99 groups), age and sex (DISCONTINUED). \url{www.statbank.dk/IND01}. Accessed 2024-12-14.}& \%\\
\midrule
0-14 years & 72,672 & 1,226,928 & 5.92\% & 968,670 & 180,346 & 18.62\%\\
15-44 years & 369,942 & 3,318,793 & 11.15\% & 2,133,883 &300,533 & 14.08\%\\
45-64 years & 396,852& 2,447,948 &16.21\% & 1,497,949 &312,122&20.84\%\\
65+ years & 681,476 & 1,582,480 & 43.06\% & 1,026,734 & 570,181&55.53\%\\
\midrule
Total & 1,520,942 & 8,576,149 & 17.75\% & 5,627,235 & 1,262,202&24.23\%\\
\
\end{tabular}
\caption{Hospital admission and population statistics for Denmark and Austria for the year 2014 }
\label{tab:pop_stats}
\end{table}

\FloatBarrier
\section{Diagnoses in the sample}
\label{app:diagnoses}
\begin{figure}
   \captionsetup{aboveskip=0pt, belowskip=15pt} 

        \centering
  \begin{subfigure}{0.7\textwidth}
    \includegraphics[width=\linewidth]{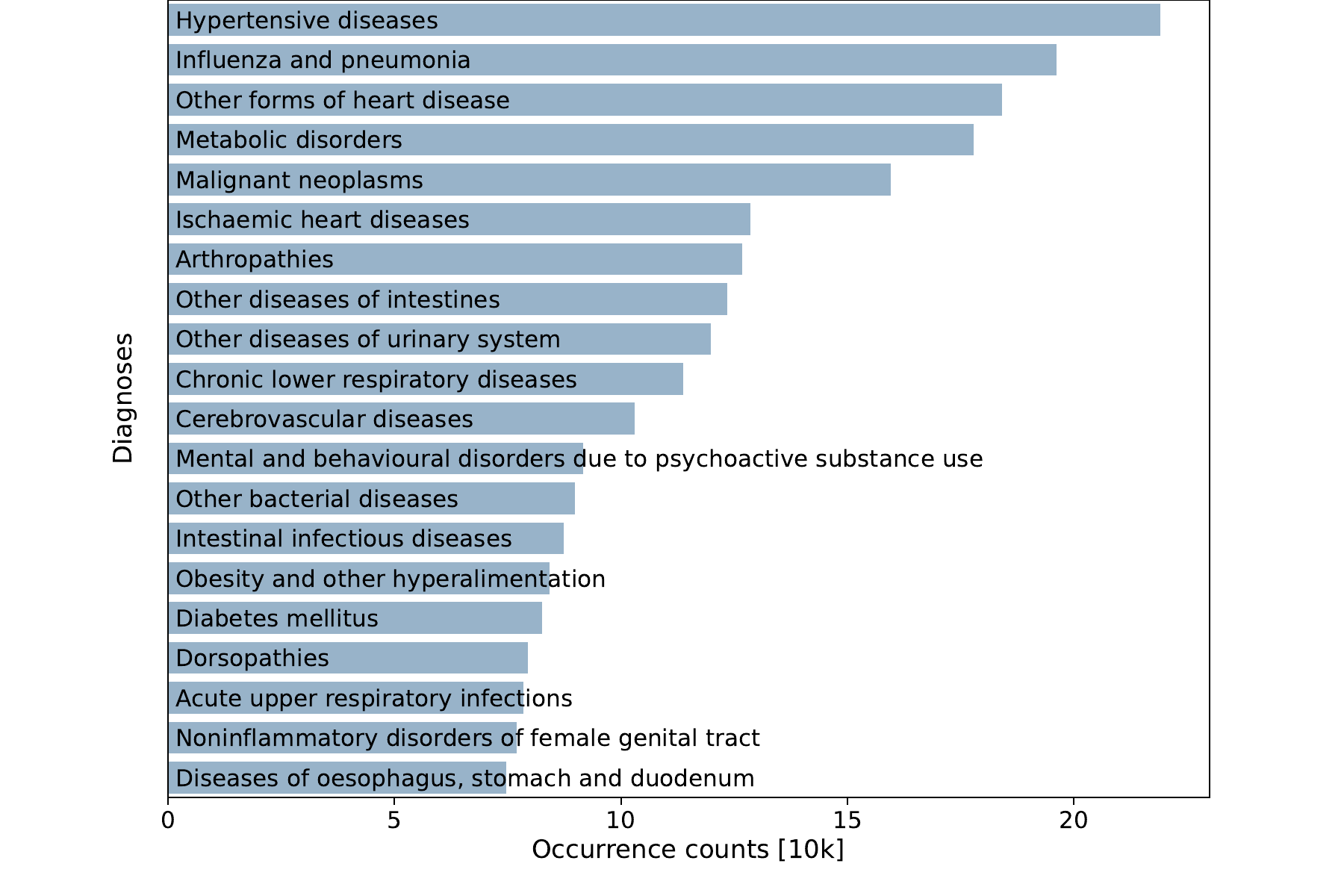}
\caption{Denmark}
    \label{fig:most_common_diag_DK}
    \end{subfigure}
    \begin{subfigure}{0.7\textwidth}

        \includegraphics[width=\linewidth]{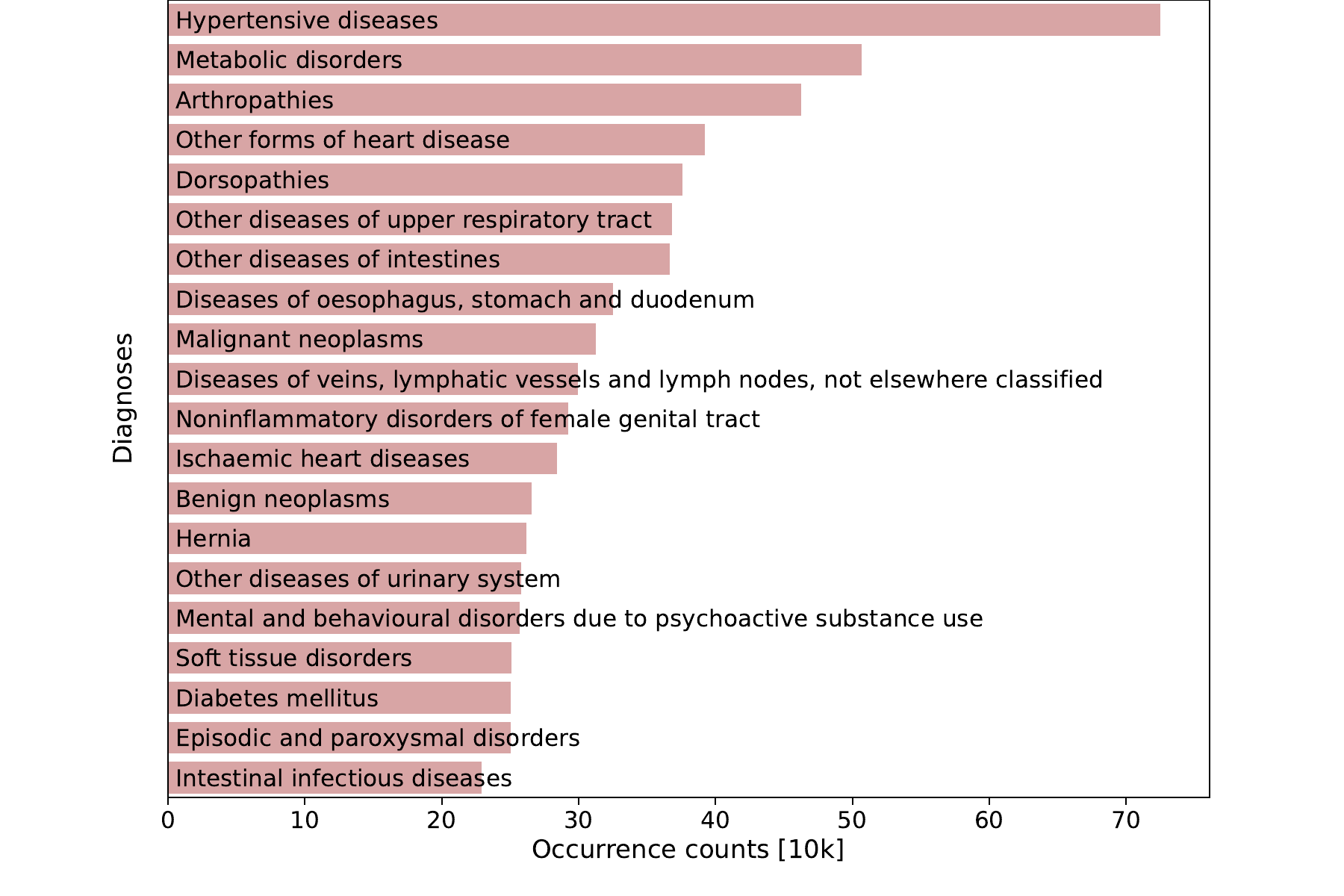}
        \caption{Austria}
        \label{fig:most_common_diag_AT}
    \end{subfigure}

    \begin{subfigure}{0.7\textwidth}

        \includegraphics[width=\linewidth]{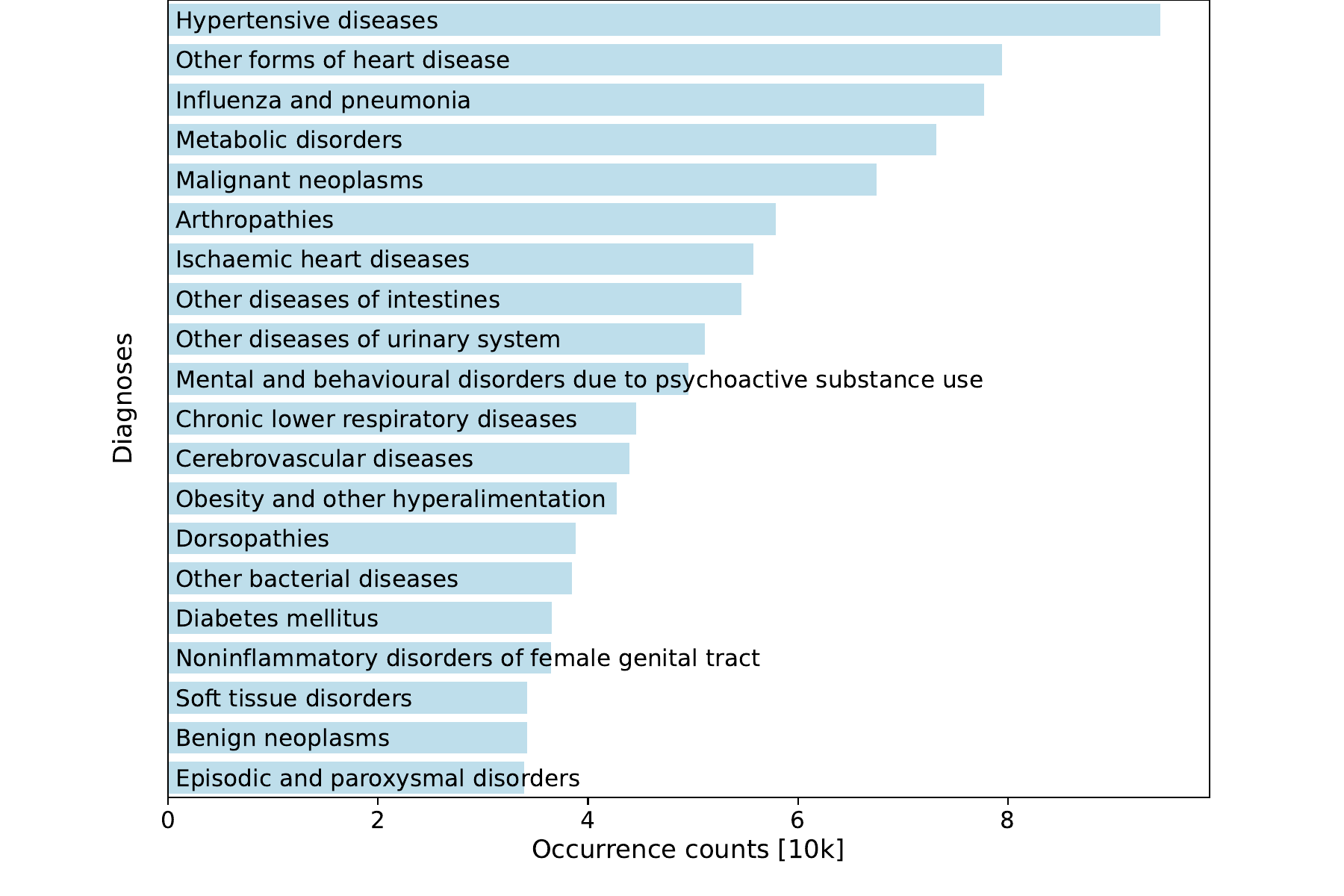}
        \caption{Denmark - matched}
        \label{fig:most_common_diag_DK_adj}
    \end{subfigure}
    \caption{20 most frequent diagnoses in the Austrian, Danish and matched Danish cohorts. Diagnoses are only counted once per individual.}
   \label{fig:most_common_diag}
\end{figure}

\begin{figure}
   \captionsetup{aboveskip=0pt, belowskip=15pt} 

        \centering
  \begin{subfigure}{0.7\textwidth}
    \includegraphics[width=\linewidth]{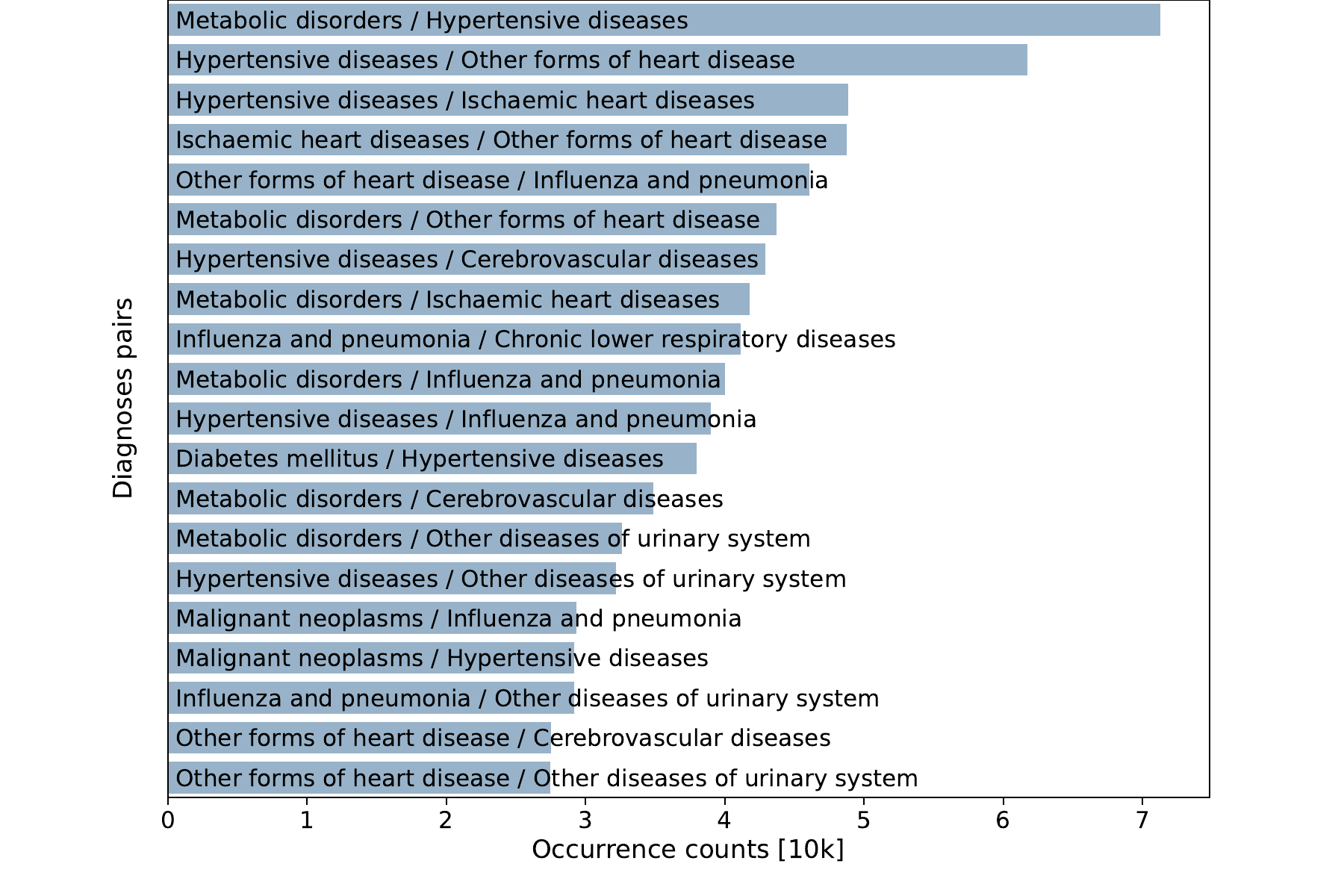}
\caption{Denmark}
    \label{fig:most_common_diag_pairs_DK}
    \end{subfigure}
   
    \begin{subfigure}{0.7\textwidth}

        \includegraphics[width=\linewidth]{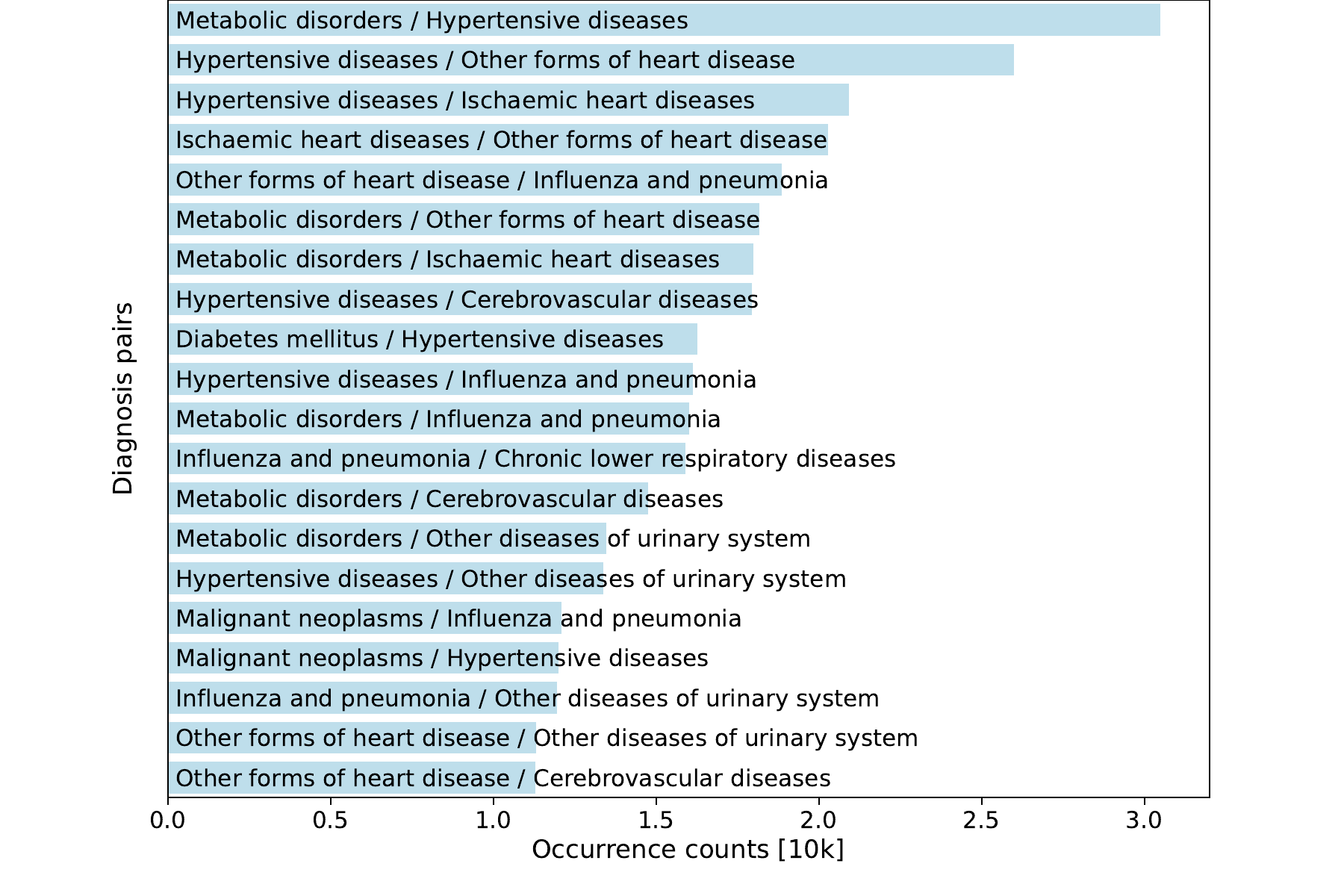}
        \caption{Denmark - matched}
        \label{fig:most_common_diag_pairs_DK_adj}
    \end{subfigure}
    \caption{20 most frequent diagnosis pairs in the Danish and matched Danish cohorts. Diagnoses are only counted once per individual.}
   \label{fig:most_common_diag}
\end{figure}

When it comes to the number of diagnoses per individual within the observation period, we find that there are less diagnoses per person in Denmark than in Austria. Overall we observe an average of 2.56 diagnoses per person in the Austrian sample, compared to 1.27 in Denmark. Figures \ref{fig:most_common_diag_DK}, \ref{fig:most_common_diag_AT} and \ref{fig:most_common_diag_DK_adj} show the occurrence of the most common diagnoses in the Danish sample, the Austrian sample and the matched Danish sample, respectively. In Figures \ref{fig:most_common_diag_pairs_DK} and \ref{fig:most_common_diag_pairs_DK_adj} we also report the most frequently occurring diagnosis pairs in the Danish and the matched Danish sample.

\begin{figure}
  \begin{subfigure}{0.9\textwidth}
 \centering
    \includegraphics[width=\linewidth]{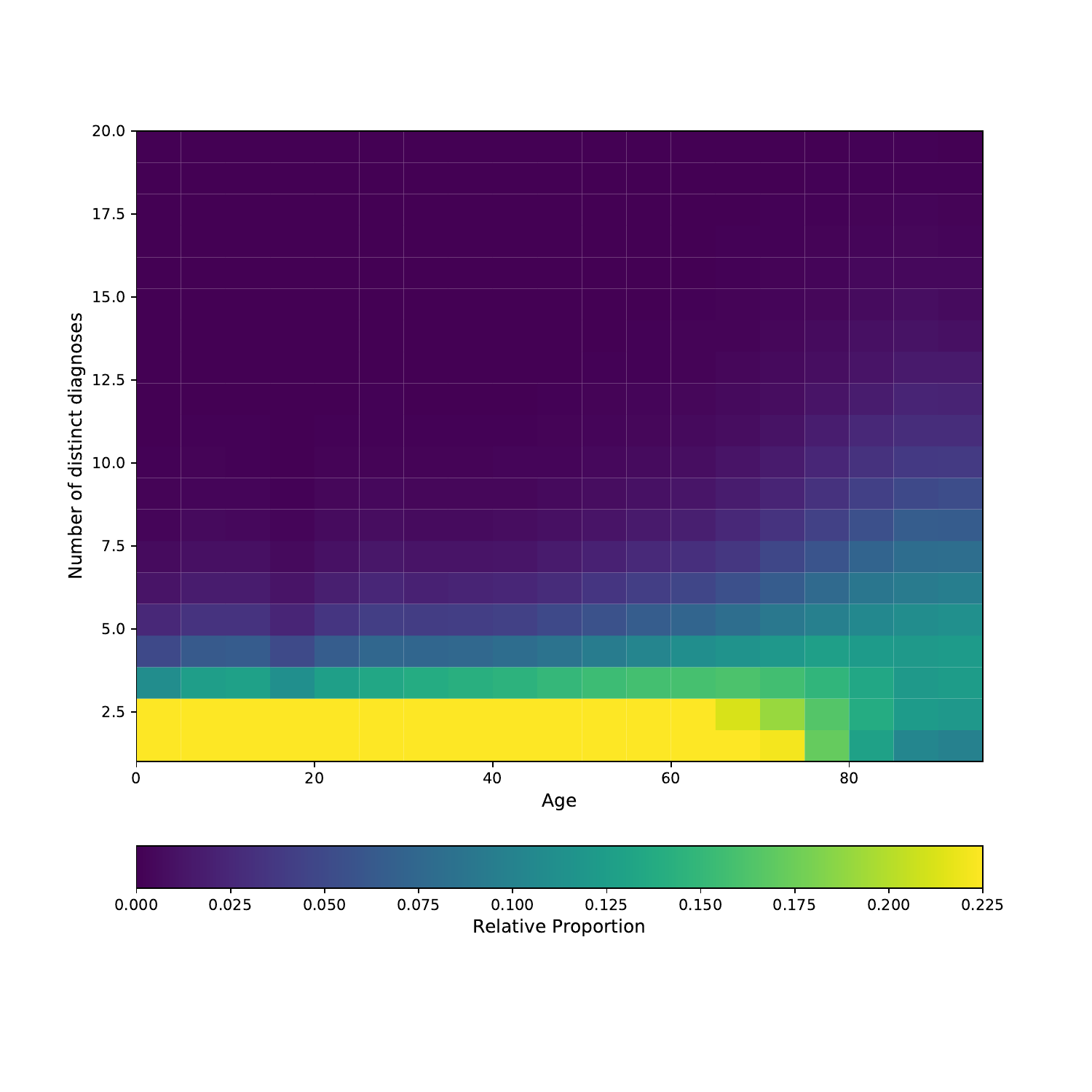}
\caption{Denmark}
    \label{fig:dist_age_diag_DK}

    \end{subfigure}

    \caption{Density of the count of unique  diagnoses acquired by patients of the study cohort during the observation period from 2003 to 2014, based on the 5-year age group they belong to in 2014. Data is normalized such
that values sum to one separately for each age group.}
   \label{fig:dist_age_diag}
\end{figure}

Figure \ref{fig:dist_age_diag} further shows the histogram of distinct diagnoses of patients in the study cohort, based on the 5-year age group they belonged to at the end of the observation period.
\FloatBarrier
\subsection{Density of diagnoses per age cohort}
\label{app_B:density}

\FloatBarrier

\section{Replication of Austrian clusters}
\label{app:replication}
\begin{figure}
   \captionsetup{aboveskip=0pt, belowskip=15pt} 

        \centering
  \includegraphics[width=0.45\textwidth]{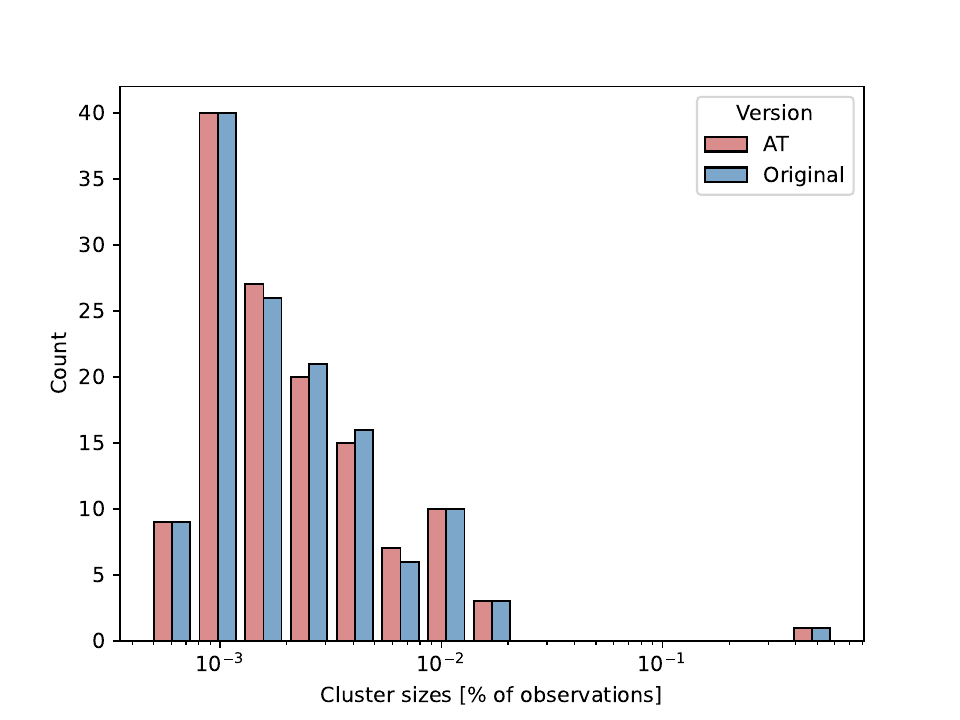}
    \caption{Distribution of cluster sizes in the original clustering on Austrian data and the replication.}
   \label{fig:cluster_sizes_AT}
\end{figure}

In a first step, we attempted a replication of the clusters produced by \cite{Haug2020}. While our results aren't fully identical they closely match them. Of the 132 clusters, 130 have identical inclusion criteria. Only clusters 73 (with \emph{Hernia} and \emph{Other diseases of the intestines} as inclusion criteria) and 78 (with \emph{Intenstinal infectious diseaeases} and \emph{Other acute lower respiratory infections} as inclusion criteria) of the original clustering couldn't be matched to a cluster with the same inclusion criteria in the replication. Figure \ref{fig:cluster_sizes_AT} shows the distribution of cluster sizes in the original and the replication, revealing only marginal discrepancies. This is also reflected in the cluster similarity measures: The NMI between the orginial and the replication is 0.995 and the ARI is 0.999.

\FloatBarrier
\section{Clustering statistics}
\label{app_c}

\begin{figure}
\centering
\includegraphics[width=0.45\textwidth]{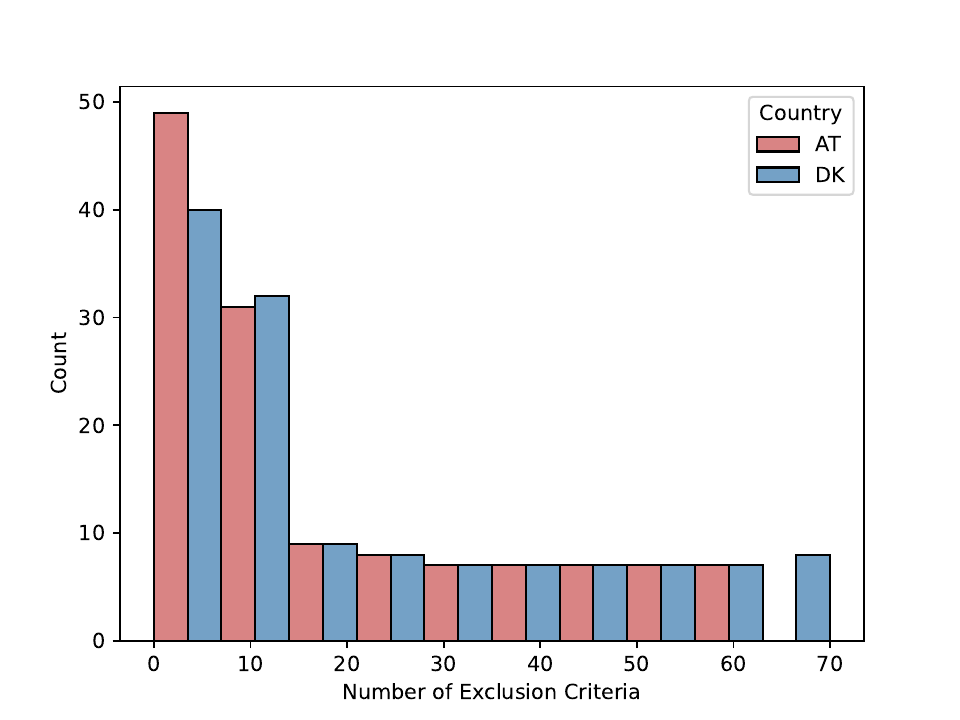}
\label{fig:criteria_dist}
\includegraphics[width=0.45\textwidth]{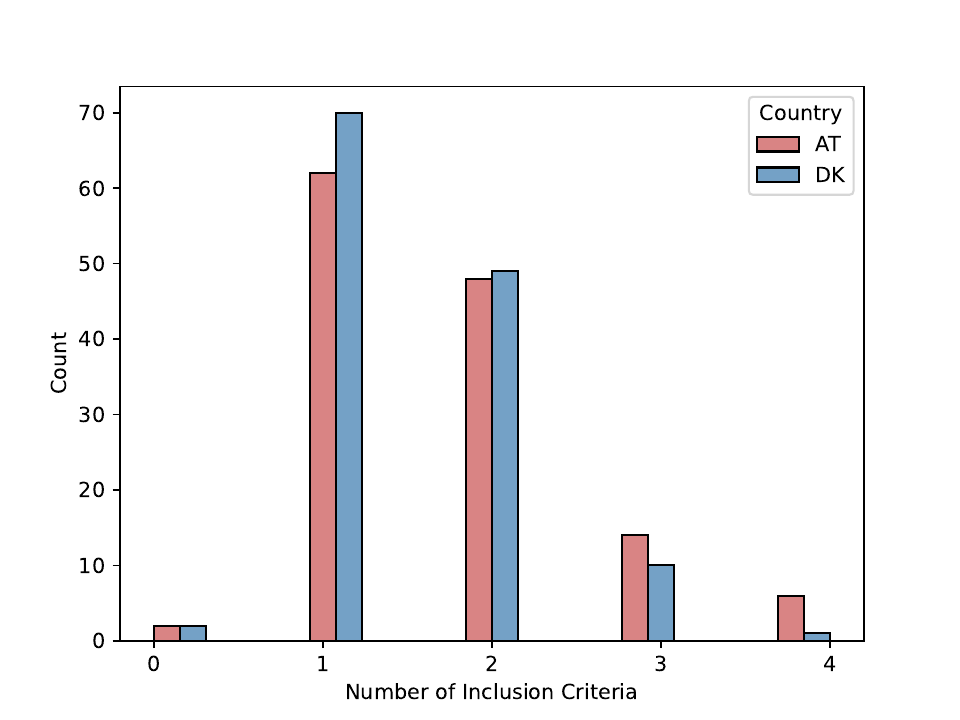}
\caption{Histogram of the observed numbers of exclusion and inclusion criteria in the Austrian and Danish clustering}
\label{fig:criteria_dist}
\end{figure}

\begin{figure}
   \captionsetup{aboveskip=0pt, belowskip=15pt} 

        \centering
  %\begin{subfigure}{\textwidth}
    \includegraphics[width=\linewidth]{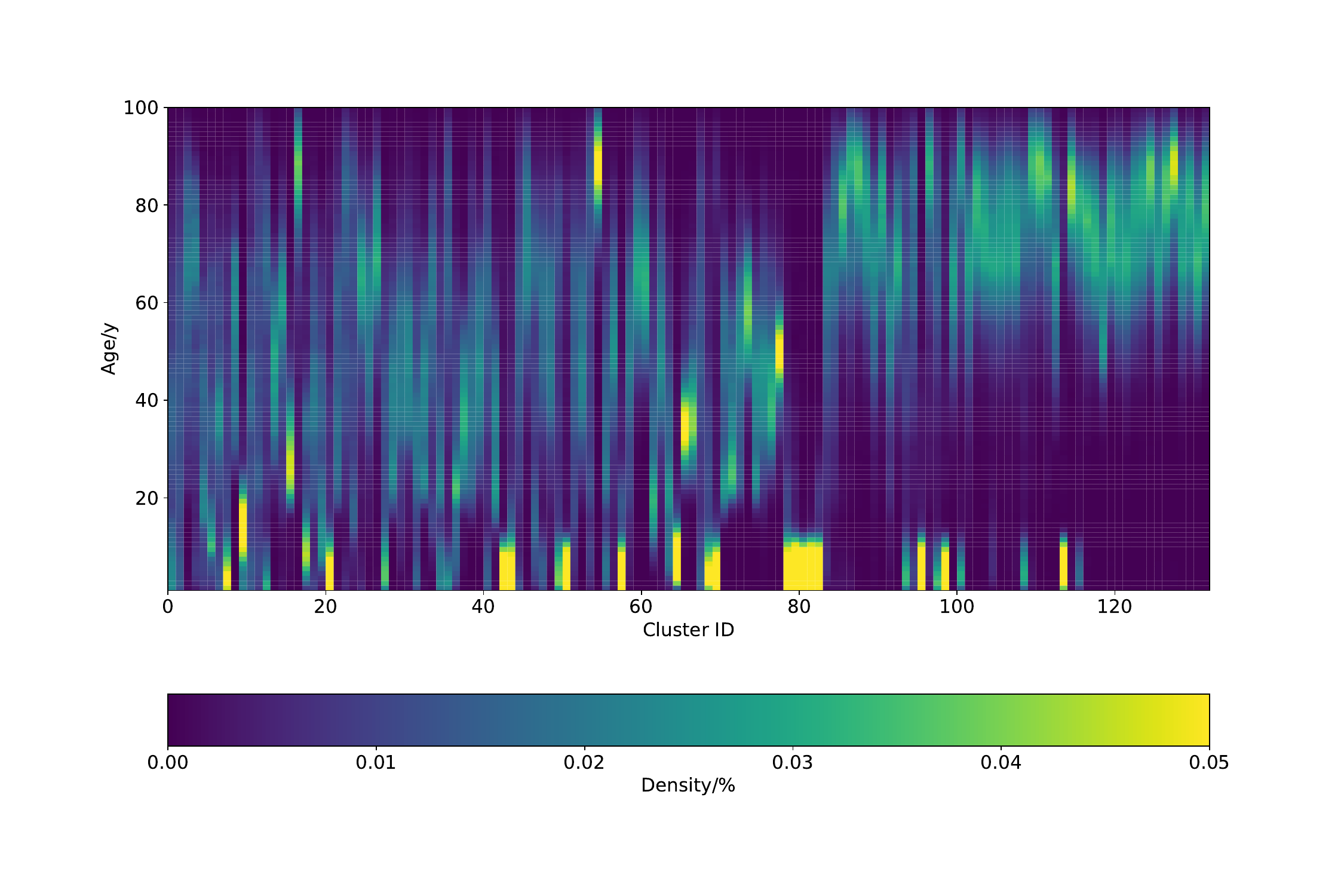}
%\caption{Denmark}
 %   \label{fig:cluster_age_density_DK}
  %  \end{subfigure}
   
    % \begin{subfigure}{\textwidth}

   %      \includegraphics[width=\linewidth]{figures/appendix_figures/03_clusters/AT_cluster_age_density.png}
   %      \caption{Austria}
   %      \label{fig:cluster_age_density_AT}
   %  \end{subfigure}
   %  \caption{Age distribution across clusters in Denmark and Austria. Note that the ordering of the clusters is not relevant.}
      \caption{Age distribution across clusters in the Danish clustering. Note that the ordering of the clusters is not relevant.}
   % \label{fig:cluster_age_density}
\end{figure}

\begin{figure}
   \captionsetup{aboveskip=0pt, belowskip=15pt} 

        \centering
  %\begin{subfigure}{\textwidth}
    \includegraphics[width=\linewidth]{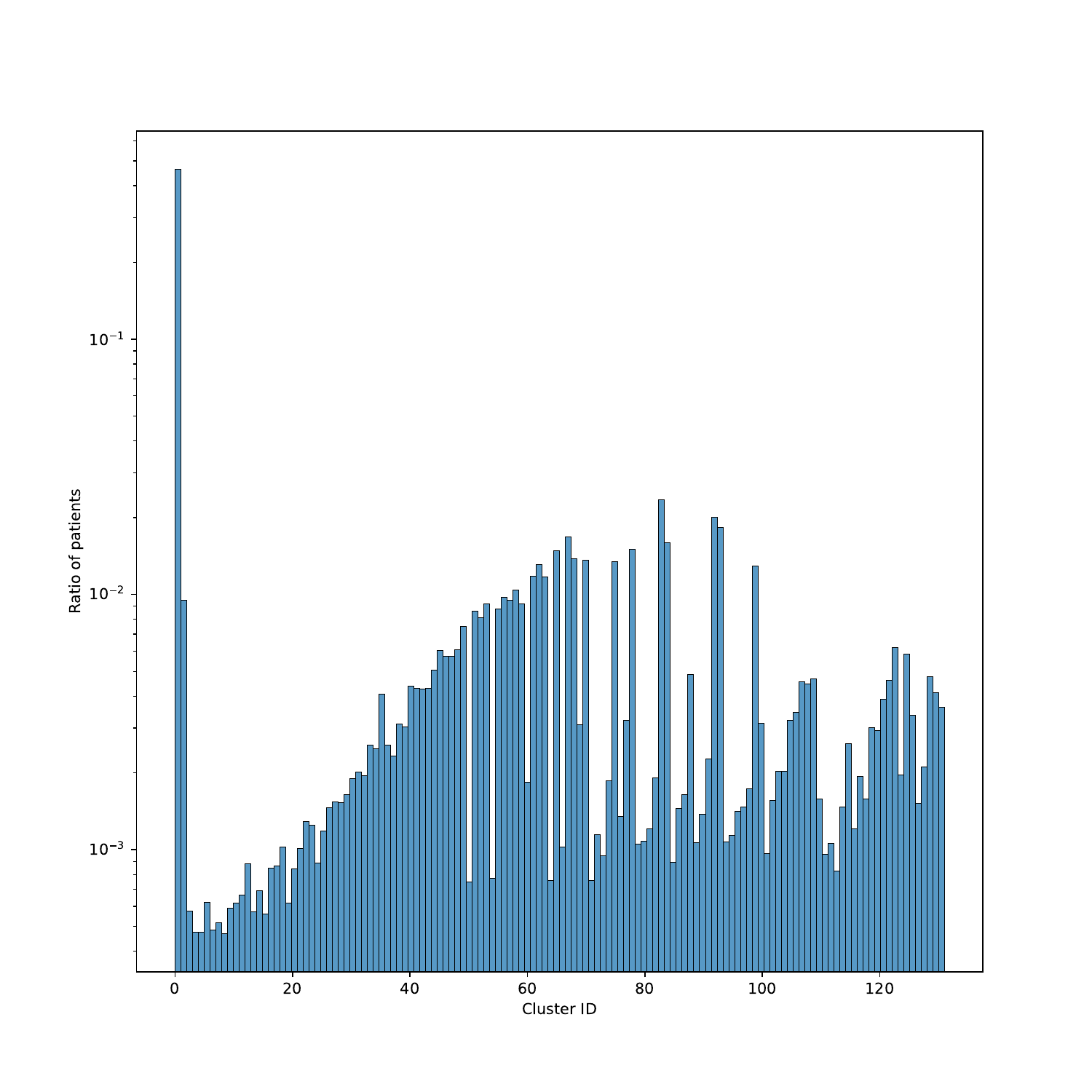}
\caption{Denmark}
    \label{fig:cluster_distribution_end_DK}

    \caption{Ratio of patients in each cluster at the end of the observation period.}
   \label{fig:cluster_distribution_end}
\end{figure}

\begin{figure}
   \captionsetup{aboveskip=0pt, belowskip=15pt} 

        \centering
  %\begin{subfigure}{\textwidth}
    \includegraphics[width=\linewidth]{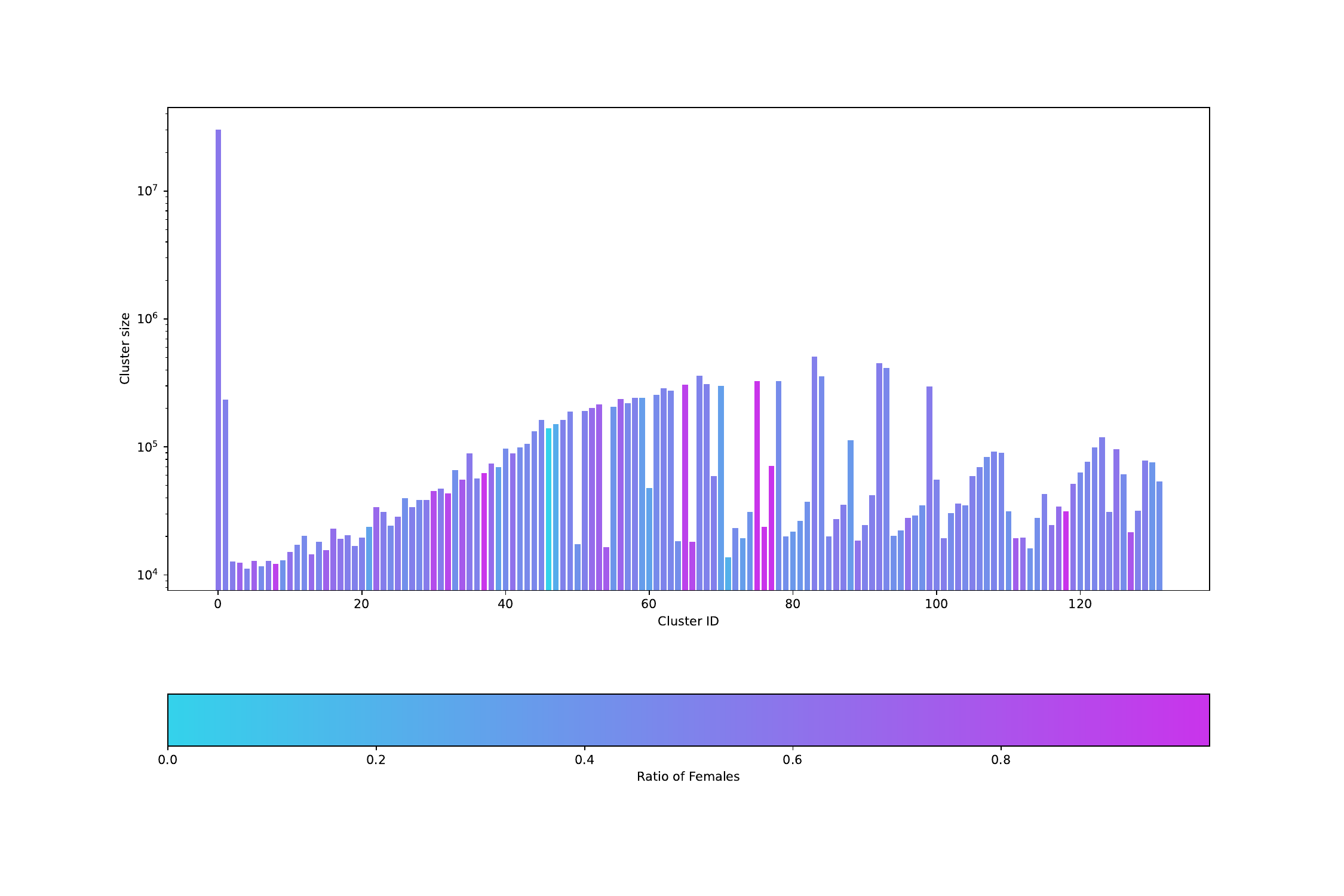}
%\caption{Denmark}
    \label{fig:cluster_gender_count_DK}

    \caption{Number of observations per cluster. Colors represent the female ratio within a cluster.}
   \label{fig:cluster_gender_count}
\end{figure}

\FloatBarrier
\section{Cross-country comparison for selected disease blocks}
\label{app:selected-diseases}

\subsection{Diabetes mellitus (E10-E14)}
In total, 82,602 patients (2.6\%) of the Danish population were diagnosed with \emph{Diabetes mellitus}. Looking at the Austrian cluster structure the most frequent two step trajectories lead from the all healthy state to cluster 6 with only \emph{Diabetes mellitus} as prerequisite and cluster 112, which has \emph{Hypertensive diseases} as an additional criterion. The most frequent 3-step trajectory (0, 102, 112) further indicates that \emph{Hypertensive diseases} often precedes diabetes while the second most frequent one (0,112, 120) indicates the frequent comorbidity with also \emph{Metabolic disorders}. 
Looking at the Danish cluster structure we observe the exact same pattern in terms of inclusion criteria, highlighting that the clustering mechanism is able to robustly capture such well understood co-morbidities.

Almost half of the individuals with a diabete diagnosis  have a reduced trajectory of length 2 in the Danish cluster structure, i.e. they move from the zero cluster to some other cluster and stay there. The most frequent trajectory (627 individuals) of length 3 goes from cluster 0 (i.e. no diagnoses) to cluster 99 which has \emph{Hypertensive diseases} as an inclusion criterion but \emph{Diabetes mellitus} still as an exclusion criterion and ends in cluster 121 with both, \emph{Hypertensive diseases} and \emph{Diabetes mellitus} as inclusion criteria.

Looking at the assignment of the Danish population with \emph{Diabetes mellitus} to the Austrian cluster structure, we also find mostly reduced trajectories of length 2. The most frequent length 3 trajectory (686), includes clusters with the same inclusion criteria as for the Danish case, showing, that both cluster structures capture these trajectories in similar ways.

\subsection{Ischaemic heart diseases (I20-I25)}
128,529 patients (4.0\%) received a diagnosis from blocks I20 to I25. The most frequent trajectories of length 3 in the Austrian cluster structure are (0, 47, 48) and (0,9,48). Cluster 9 only has \emph{Ischaemic heart diseases} as inclusion criterion, while cluster 47 has only \emph{Other forms of heart disease} and cluster 48 combines both. 
In the Danish cluster structure, the two most frequent trajectories of length 3 are identical to the ones from the Austrian data when focusing only on the inclusion criteria of the clusters, indicating a strong robustness of the clustering mechanism with respect to capturing these comorbidities.

\subsection{Chronic lower respiratory diseases (J40-J47)}
113,786 individuals (3.5\%) in the population have at some point been diagnosed with \emph{Chronic lower respiratory diseases}. Apart from the most frequent (22,575) two-step trajectory, with patients going from a healthy state to cluster 28 with 
\emph{Chronic lower respiratory diseases} as the single inclusion criterion, 7,128 go directly from the zero cluster to cluster 37 and an additional 1,448 transition from cluster 28 to cluster 37, a cluster with only \emph{Influenza and pneumonia} as inclusion requirement.
Similarly, looking at the Danish cluster structure we find the two most frequent 3 step trajectories ending in cluster 108  which has both \emph{Influenza pneumonia} and \emph{Chronic lower respiratory diseases} as inclusion criteria. Going from healthy to cluster 108 is also the second most frequent two step trajectory after (0, 49) which only involves \emph{Chronic lower respiratory diseases}. The existence of a frequent Danish cluster that also requires \emph{Influenza pneumonia}, might be an indication that this disease is more closely linked to \emph{Chronic lwoer respiratory diseases} in Denmark as compared to Austria.

\subsection{Cerebrovascular diseases (I60-I69)}
103,049 patients (3.18\%) in the Danish population were diagnosed with \emph{Cerebrovascular diseases}. When assigning them to the Austrian cluster structure, we find that around half of them have 2 step trajectories, with the most frequent one going from cluster 0 to cluster 11, which has \emph{Cerebrovascular diseases} as sole inclusion criterion and 51 exclusion criteria. The most frequent trajectory of length 3 is (0,102,107), leading patients from no diagnosis, to \emph{Hypertensive diseases} to then also acquiring \emph{Cerebrovascular diseases}. This is similar to the most frequent trajectory observed for the Austrian population (\cite{Haug2020}, p.5), although patients with a 3 step trajectory are most likely to be diagnosed with \emph{Hypertensive diseases} simultaneously with \emph{Other forms of heart diseases} before also being diagnosed with \emph{Cerebrovascular disease} (Trajectory 0-114-123).
If we assign patients to the Danish cluster structure and observe the most frequent 3-step trajectories, we find that the 3 most frequent ones all involve cluster 122 that has not only \emph{Cerebrovascular disease} as an inclusion criterion but also \emph{Hypertensive disease}. Generally speaking, patterns are thus similar as in the Austrian structure.

\subsection{Malignant neoplasms (C00-C97)}
A total of 159,470 patients (4.9\%) in the Danish population were diagnosed with \emph{Malignant neoplasms}. Assigned to the Austrian clusters, 68,487 (43.2\%) of patients go directly from the healthy state to cluster 55 with \emph{Malignent neoplasms} as sole inclusion criterion and remain there for the time of the observation period. Other frequent two step trajectories lead from the 0 cluster to cluster 74 (4,375; 2.8\% ;\emph{Malignent neoplasms, Other diseases of intestines}) and cluster 109 (4,242; 2.7\%; \emph{Malignent neoplasms, Hypertensive diseases}). The most frequent reduced trajectories with 3 distinct health states are (0,55,74), followed by 2,454 (1.5\%) patients and (0,1,55), followed by 2,443 patients. Cluster 1 is a cluster with 62 exclusion criteria and no inclusion criteria, indicating that those individuals have been diagnosed with some disease (otherwise they would be in the zero cluster) but don't show a particularly frequent morbidity pattern. In the Austrian population the trajectory (0, 55) is similarly the most frequent one.
% If we look at the Danish cluster structure, we observe 53,006 patients with this diagnosis (33.2\%) going from cluster 0 to cluster 92 with \emph{Malignant neoplasms} as sole inclusion criterion. The three most frequent 3 step trajectories further append clusters to this trajectory with each one additional inclusion criterion - namely 107 (\emph{Influenza pneumonia}), 105 (\emph{Other diseases of intestines}) and 106 (\emph{Aplastic and other anameias}).

\subsection{Mood (affective) disorders (F30-F39)}
A total of 48,993 (1.5\%) of the Danish population were diagnosed with a mood disorder in a hospital during the observation period. This is comparatively less than in the Austrian population, where 4.1\% received this diagnosis. 
When assigned to the Austrian cluster structure 10,076 (20.5\%) of the Danish patients with this diagnosis go from the healthy cluster to cluster 27 with just \emph{Mood affective disorders} as an inclusion criterion. Around 5\% each go from cluster 0 to clusters 71 with both \emph{Mood affective disorders} and  \emph{Mental and behavioural disorders due to psychoactive substance abuse} as inclusion criterion; and cluster 41 which has as only inclusion criterion \emph{Neurotic, stress-related and somatoform disorders}. The most frequent 3 step trajectory is the same as for the Austrian population, specifically (0, 64, 71), indicating that in both countries, \emph{Mental and behavioural disorders due to psychoactive substance abuse} often preceed or are simultaneously diagnosed with \emph{Mood affective disorders}.

In the Danish cluster structure 9,427 (19\%) of individuals with this diagnosis go from healthy to cluster 38 which--similarly to Austrian cluster 27--- only has \emph{Mood disorders} as inclusion criterion. The most frequent 3 step trajectory is from cluster 0 to cluster 70 with \emph{Mental and behavioural disorders due to psychoactive substance use} as inclusion criterion and subsequently to cluster 72 which adds \emph{Mood disorders}.

\subsection{Low overlap clusters}
Description of the Danish clusters with the lowest overlap with a single Austrian cluster.
\textbf{Cluster 100.}

Cluster 100 has \emph{Metabolic disorders} as an inclusion criterion. The largest share of its observations would be assigned to the Austrian cluster 40, with \emph{Renal tubulo-interstitial diseases} and \emph{Urolithiasis} as inclusion criteria. The second largest share (14\%) is in Austrian cluster 75 with inclusion criteria \emph{Other diseases of intestines} and \emph{Diseases of veins, lymphatic vessels and lymph nodes, not elsewhere classified}. Another 13\% are in cluster 41, requiring observations to be diagnosed with \emph{Neurotic, stress-related and somatoform disorders}.

\textbf{Cluster 113.}
The Danish cluster 113 has \emph{Influenza pneumonia}, \emph{Other acute lower respiratory infections} and \emph{Chronic lower respiratory diseases} as inclusion criteria. Like Cluster 100, the largest share of its observations are within the Austrian Cluster 40. Another, 20\% of this cluster's population is assigned to cluster 55 in the Austrian clustering scheme. This cluster has \emph{Acute upper respiratory infections} and \emph{
Other acute lower respiratory infections} as inclusion criteria.

\textbf{Cluster 114.}
The Danish cluster with number 114 has three inclusion criteria: \emph{Other forms of heart disease}, \emph{Influenza and pneumonia} and \emph{Chronic lower respiratory diseases}. It further has two exclusion criteria: \emph{Hypertensive disease} and \emph{Other diseases of the respiratory system}, differentiating it from other clusters that include chronic lower respiratory disease. 33\% of observations in this cluster would be in the Austrian cluster 48 which similarly includes \emph{Other forms of heart disease} but additionally requires \emph{Ischaemic heart diseases} for inclusion. An other 15\% are allocated to Austrian cluster 49, indicating the these individuals have also been diagnosed with \emph{Episodic and paroxysmal disorders}.

\textbf{Cluster 115.}
Cluster 115 is similarly characterized by \emph{Influenza pneumonia},\emph{Chronic lower respiratory diseases} and  \emph{Other disease of the lower respiratory system}, with the single exclusion criterion of \emph{Hypertensive disease}. The largest shares of these observations  are assigned to the earlier described Austrian clusters 40  (30\%) and 48 (11\%). 

\section{Cluster specifications}
\label{app:cluster_criteria}
\newcolumntype{P}[1]{>{\raggedright\arraybackslash}p{#1}}
\renewcommand{\arraystretch}{0.8}
\input{tables/tables_appendix/Cluster_1}

\input{tables/tables_appendix/Cluster_2}

\input{tables/tables_appendix/Cluster_3}
\input{tables/tables_appendix/Cluster_4}
\input{tables/tables_appendix/Cluster_5}
\input{tables/tables_appendix/Cluster_6}
\input{tables/tables_appendix/Cluster_7}
\input{tables/tables_appendix/Cluster_8}
\input{tables/tables_appendix/Cluster_9}
\input{tables/tables_appendix/Cluster_10}
\input{tables/tables_appendix/Cluster_11}
\input{tables/tables_appendix/Cluster_12}
\input{tables/tables_appendix/Cluster_13}
\input{tables/tables_appendix/Cluster_14}
\input{tables/tables_appendix/Cluster_15}
\FloatBarrier
\input{tables/tables_appendix/Cluster_16}
\FloatBarrier
\input{tables/tables_appendix/Cluster_17}
\FloatBarrier
\input{tables/tables_appendix/Cluster_18}
\FloatBarrier
\input{tables/tables_appendix/Cluster_19}
\FloatBarrier
\input{tables/tables_appendix/Cluster_20}
\FloatBarrier
\input{tables/tables_appendix/Cluster_21}
\FloatBarrier
\input{tables/tables_appendix/Cluster_22}
\FloatBarrier
\input{tables/tables_appendix/Cluster_23}
\FloatBarrier
\input{tables/tables_appendix/Cluster_24}
\FloatBarrier
\input{tables/tables_appendix/Cluster_25}
\FloatBarrier
\input{tables/tables_appendix/Cluster_26}
\FloatBarrier
\input{tables/tables_appendix/Cluster_27}
\FloatBarrier
\input{tables/tables_appendix/Cluster_28}
\FloatBarrier
\input{tables/tables_appendix/Cluster_29}
\FloatBarrier
\input{tables/tables_appendix/Cluster_30}
\FloatBarrier
\input{tables/tables_appendix/Cluster_31}
\FloatBarrier
\input{tables/tables_appendix/Cluster_32}
\FloatBarrier
\input{tables/tables_appendix/Cluster_33}
\FloatBarrier
\input{tables/tables_appendix/Cluster_34}
\FloatBarrier
\input{tables/tables_appendix/Cluster_35}
\FloatBarrier
\input{tables/tables_appendix/Cluster_36}
\FloatBarrier
\input{tables/tables_appendix/Cluster_37}
\FloatBarrier
\input{tables/tables_appendix/Cluster_38}
\FloatBarrier
\input{tables/tables_appendix/Cluster_39}
\FloatBarrier
\input{tables/tables_appendix/Cluster_40}
\FloatBarrier
\input{tables/tables_appendix/Cluster_41}
\FloatBarrier
\input{tables/tables_appendix/Cluster_42}
\FloatBarrier
\input{tables/tables_appendix/Cluster_43}
\FloatBarrier
\input{tables/tables_appendix/Cluster_44}
\FloatBarrier
\input{tables/tables_appendix/Cluster_45}
\FloatBarrier
\input{tables/tables_appendix/Cluster_46}
\FloatBarrier
\input{tables/tables_appendix/Cluster_47}
\FloatBarrier
\input{tables/tables_appendix/Cluster_48}
\FloatBarrier
\input{tables/tables_appendix/Cluster_49}
\FloatBarrier
\input{tables/tables_appendix/Cluster_50}
\FloatBarrier
\input{tables/tables_appendix/Cluster_51}
\FloatBarrier
\input{tables/tables_appendix/Cluster_52}
\FloatBarrier
\input{tables/tables_appendix/Cluster_53}
\FloatBarrier
\input{tables/tables_appendix/Cluster_54}
\FloatBarrier
\input{tables/tables_appendix/Cluster_55}
\FloatBarrier
\input{tables/tables_appendix/Cluster_56}
\FloatBarrier
\input{tables/tables_appendix/Cluster_57}
\FloatBarrier
\input{tables/tables_appendix/Cluster_58}
\FloatBarrier
\input{tables/tables_appendix/Cluster_59}
\FloatBarrier
\input{tables/tables_appendix/Cluster_60}
\FloatBarrier
\input{tables/tables_appendix/Cluster_61}
\FloatBarrier
\input{tables/tables_appendix/Cluster_62}
\FloatBarrier
\input{tables/tables_appendix/Cluster_63}
\FloatBarrier
\input{tables/tables_appendix/Cluster_64}
\FloatBarrier
\input{tables/tables_appendix/Cluster_65}
\FloatBarrier
\input{tables/tables_appendix/Cluster_66}
\FloatBarrier
\input{tables/tables_appendix/Cluster_67}
\FloatBarrier
\input{tables/tables_appendix/Cluster_68}
\FloatBarrier
\input{tables/tables_appendix/Cluster_69}
\FloatBarrier
\input{tables/tables_appendix/Cluster_70}
\FloatBarrier
\input{tables/tables_appendix/Cluster_71}
\FloatBarrier
\input{tables/tables_appendix/Cluster_72}
\FloatBarrier
\input{tables/tables_appendix/Cluster_73}
\FloatBarrier
\input{tables/tables_appendix/Cluster_74}
\FloatBarrier
\input{tables/tables_appendix/Cluster_75}
\FloatBarrier
\input{tables/tables_appendix/Cluster_76}
\FloatBarrier
\input{tables/tables_appendix/Cluster_77}
\FloatBarrier
\input{tables/tables_appendix/Cluster_78}
\FloatBarrier
\input{tables/tables_appendix/Cluster_79}
\FloatBarrier
\input{tables/tables_appendix/Cluster_80}
\FloatBarrier
\input{tables/tables_appendix/Cluster_81}
\FloatBarrier
\input{tables/tables_appendix/Cluster_82}
\FloatBarrier
\input{tables/tables_appendix/Cluster_83}
\FloatBarrier
\input{tables/tables_appendix/Cluster_84}
\FloatBarrier
\input{tables/tables_appendix/Cluster_85}
\FloatBarrier
\input{tables/tables_appendix/Cluster_86}
\FloatBarrier
\input{tables/tables_appendix/Cluster_87}
\FloatBarrier
\input{tables/tables_appendix/Cluster_88}
\FloatBarrier
\input{tables/tables_appendix/Cluster_89}
\FloatBarrier
\input{tables/tables_appendix/Cluster_90}
\FloatBarrier
\input{tables/tables_appendix/Cluster_91}
\FloatBarrier
\input{tables/tables_appendix/Cluster_92}
\FloatBarrier
\input{tables/tables_appendix/Cluster_93}
\FloatBarrier
\input{tables/tables_appendix/Cluster_94}
\FloatBarrier
\input{tables/tables_appendix/Cluster_95}
\FloatBarrier
\input{tables/tables_appendix/Cluster_96}
\FloatBarrier
\input{tables/tables_appendix/Cluster_97}
\FloatBarrier
\input{tables/tables_appendix/Cluster_98}
\FloatBarrier
\input{tables/tables_appendix/Cluster_99}
\FloatBarrier
\input{tables/tables_appendix/Cluster_100}
\FloatBarrier
\input{tables/tables_appendix/Cluster_101}
\FloatBarrier
\input{tables/tables_appendix/Cluster_102}
\FloatBarrier
\input{tables/tables_appendix/Cluster_103}
\FloatBarrier
\input{tables/tables_appendix/Cluster_104}
\FloatBarrier
\input{tables/tables_appendix/Cluster_105}
\FloatBarrier
\input{tables/tables_appendix/Cluster_106}
\FloatBarrier
\input{tables/tables_appendix/Cluster_107}
\FloatBarrier
\input{tables/tables_appendix/Cluster_108}
\FloatBarrier
\input{tables/tables_appendix/Cluster_109}
\FloatBarrier
\input{tables/tables_appendix/Cluster_110}
\FloatBarrier
\input{tables/tables_appendix/Cluster_111}
\FloatBarrier
\input{tables/tables_appendix/Cluster_112}
\FloatBarrier
\input{tables/tables_appendix/Cluster_113}
\FloatBarrier
\input{tables/tables_appendix/Cluster_114}
\FloatBarrier
\input{tables/tables_appendix/Cluster_115}
\FloatBarrier
\input{tables/tables_appendix/Cluster_116}
\FloatBarrier
\input{tables/tables_appendix/Cluster_117}
\FloatBarrier
\input{tables/tables_appendix/Cluster_118}
\FloatBarrier
\input{tables/tables_appendix/Cluster_119}
\FloatBarrier
\input{tables/tables_appendix/Cluster_120}
\FloatBarrier
\input{tables/tables_appendix/Cluster_121}
\FloatBarrier
\input{tables/tables_appendix/Cluster_122}
\FloatBarrier
\input{tables/tables_appendix/Cluster_123}
\FloatBarrier
\input{tables/tables_appendix/Cluster_124}
\FloatBarrier
\input{tables/tables_appendix/Cluster_125}
\FloatBarrier
\input{tables/tables_appendix/Cluster_126}
\FloatBarrier
\input{tables/tables_appendix/Cluster_127}
\FloatBarrier
\input{tables/tables_appendix/Cluster_128}
\FloatBarrier
\input{tables/tables_appendix/Cluster_129}
\FloatBarrier
\input{tables/tables_appendix/Cluster_130}
\FloatBarrier
\input{tables/tables_appendix/Cluster_131}
\FloatBarrier

\section{Results for the matched sample}